%% file: main.tex
\documentclass[12pt,a4paper]{article}
\pdfoutput=1
\usepackage{ifthen} % for conditional statements
\newboolean{pdflatex}
\setboolean{pdflatex}{true} % False for eps figures 

\newboolean{articletitles}
\setboolean{articletitles}{true} % False removes titles in references

\newboolean{uprightparticles}
\setboolean{uprightparticles}{false} %True for upright particle symbols

\newboolean{inbibliography}
\setboolean{inbibliography}{false} %True once you enter the bibliography

\input{preamble}
\usepackage{longtable} % only for template; not usually to be used in PAPERs
\usepackage{subfigure}
\usepackage{pifont}
\begin{document}

%%%%%%%%%%%%%%%%%%%%%%%%%
%%%%% Title     %%%%%%%%%
%%%%%%%%%%%%%%%%%%%%%%%%%
\renewcommand{\thefootnote}{\fnsymbol{footnote}}
\setcounter{footnote}{1}

% %%%%%%% CHOOSE TITLE PAGE--------
%\onecolumn
%\input{title-LHCb-ANA}
%\input{title-LHCb-CONF}
\input{title-LHCb-PAPER}

%\twocolumn
% %%%%%%%%%%%%% ---------

\renewcommand{\thefootnote}{\arabic{footnote}}
\setcounter{footnote}{0}

%%%%%%%%%%%%%%%%%%%%%%%%%%%%%%%%
%%%%%  Table of Content   %%%%%%
%%%%%%%%%%%%%%%%%%%%%%%%%%%%%%%%
%%%% Uncomment next 2 lines if desired
%\tableofcontents
%\cleardoublepage

%%%%%%%%%%%%%%%%%%%%%%%%%
%%%%% Main text %%%%%%%%%
%%%%%%%%%%%%%%%%%%%%%%%%%

\pagestyle{plain} % restore page numbers for the main text
\setcounter{page}{1}
\pagenumbering{arabic}

%% Uncomment during review phase. 
%% Comment before a final submission.
%\linenumbers

% You can include short sections directly in the main tex file.
% However, for larger papers it is desirable to split the text into
% several semiautonomous files, which can be revised independently.
% This is especially useful when developing a document in
% collaboration with several people, since then different parts can be
% edited independently.  This type of file organization is shown here.
% 

\input{introduction}
\input{detector}
\input{candidateSelection}
\input{eventYield}
\input{crosssectionMeasurement}
\input{systematicUncertainties}
\input{results}
\input{conclusions}
\input{acknowledgements}
\newpage
\input{appendix}

\newpage
\addcontentsline{toc}{section}{References}
\setboolean{inbibliography}{true}
\bibliographystyle{LHCb}
\bibliography{mine,main,LHCb-PAPER,LHCb-CONF,LHCb-DP,LHCb-TDR}

\newpage
\input{LHCb_Authorship_flat_07-Jun-2016.tex}

\end{document}

%% file: preamble.tex
\usepackage[top=1in, bottom=1.25in, left=1in, right=1in]{geometry}

\usepackage[percent]{overpic}
\usepackage[normalem]{ulem}

\usepackage{microtype}
\usepackage{lineno}  % for line numbering during review
\usepackage{xspace} % To avoid problems with missing or double spaces after
\usepackage{caption} %these three command get the figure and table captions automatically small

\usepackage{floatrow}
\usepackage{graphicx}  % to include figures (can also use other packages)
\usepackage{color}
\usepackage{colortbl}
\graphicspath{{./figs/}} % Make Latex search fig subdir for figures

\usepackage{amsmath} % Adds a large collection of math symbols
\usepackage{amssymb}
\usepackage{amsfonts}
\usepackage{upgreek} % Adds in support for greek letters in roman typeset

\newcommand*\patchAmsMathEnvironmentForLineno[1]{%
\expandafter\let\csname old#1\expandafter\endcsname\csname #1\endcsname
\expandafter\let\csname oldend#1\expandafter\endcsname\csname
end#1\endcsname
 \renewenvironment{#1}%
   {\linenomath\csname old#1\endcsname}%
   {\csname oldend#1\endcsname\endlinenomath}%
}
\newcommand*\patchBothAmsMathEnvironmentsForLineno[1]{%
  \patchAmsMathEnvironmentForLineno{#1}%
  \patchAmsMathEnvironmentForLineno{#1*}%
}
\AtBeginDocument{%
\patchBothAmsMathEnvironmentsForLineno{equation}%
\patchBothAmsMathEnvironmentsForLineno{align}%
\patchBothAmsMathEnvironmentsForLineno{flalign}%
\patchBothAmsMathEnvironmentsForLineno{alignat}%
\patchBothAmsMathEnvironmentsForLineno{gather}%
\patchBothAmsMathEnvironmentsForLineno{multline}%
\patchBothAmsMathEnvironmentsForLineno{eqnarray}%
}

\usepackage{hyperref}    % Hyperlinks in references
\usepackage[all]{hypcap} % Internal hyperlinks to floats.

\input{lhcb-symbols-def} % Add in the predefined LHCb symbols

\usepackage{cite} % Allows for ranges in citations
\usepackage{mciteplus}

%% file: lhcb-symbols-def.tex
\def\myTilde {\raise.17ex\hbox{$\scriptstyle\mathtt{\sim}$}}
\def\cteq     {\mbox{\textsc{CTEQ6L1}}\xspace}
\def\mcfm     {\mbox{\textsc{MCFM}}\xspace}

\def\lhcb {\mbox{LHCb}\xspace}

\def\MagUp {\mbox{\em Mag\kern -0.05em Up}\xspace}

\ifthenelse{\boolean{uprightparticles}}%
{

 \def\Pmu         {\ensuremath{\upmu}\xspace}                 
 \def\Pnu         {\ensuremath{\upnu}\xspace}

 \def\Ptau        {\ensuremath{\uptau}\xspace}

 \def\PDelta      {\ensuremath{\Delta}\xspace}                 
 \def\PXi      {\ensuremath{\Xi}\xspace}                 
 \def\PLambda      {\ensuremath{\Lambda}\xspace}                 
 \def\PSigma      {\ensuremath{\Sigma}\xspace}                 
 \def\POmega      {\ensuremath{\Omega}\xspace}                 
 \def\PUpsilon      {\ensuremath{\Upsilon}\xspace}                 
 
 \def\PB      {\ensuremath{\mathrm{B}}\xspace}                 
                  
 \def\PD      {\ensuremath{\mathrm{D}}\xspace}

 \def\PK      {\ensuremath{\mathrm{K}}\xspace}

 \def\PW      {\ensuremath{\mathrm{W}}\xspace}

 \def\PZ      {\ensuremath{\mathrm{Z}}\xspace}                 
                  
 \def\Pb      {\ensuremath{\mathrm{b}}\xspace}                 
 \def\Pc      {\ensuremath{\mathrm{c}}\xspace}                 
                  
 \def\Pe      {\ensuremath{\mathrm{e}}\xspace}

 \def\Pi      {\ensuremath{\mathrm{i}}\xspace}

 \def\Pt      {\ensuremath{\mathrm{t}}\xspace}

}
{

 \def\Pmu         {\ensuremath{\mu}\xspace}                 
 \def\Pnu         {\ensuremath{\nu}\xspace}

 \def\Ptau        {\ensuremath{\tau}\xspace}

 \mathchardef\PDelta="7101
 \mathchardef\PXi="7104
 \mathchardef\PLambda="7103
 \mathchardef\PSigma="7106
 \mathchardef\POmega="710A
 \mathchardef\PUpsilon="7107
                  
 \def\PB      {\ensuremath{B}\xspace}                 
                  
 \def\PD      {\ensuremath{D}\xspace}

 \def\PK      {\ensuremath{K}\xspace}

 \def\PW      {\ensuremath{W}\xspace}

 \def\PZ      {\ensuremath{Z}\xspace}                 
                  
 \def\Pb      {\ensuremath{b}\xspace}                 
 \def\Pc      {\ensuremath{c}\xspace}                 
                  
 \def\Pe      {\ensuremath{e}\xspace}

 \def\Pi      {\ensuremath{i}\xspace}

 \def\Pt      {\ensuremath{t}\xspace}

}

\makeatletter
\ifcase \@ptsize \relax% 10pt
  \newcommand{\miniscule}{\@setfontsize\miniscule{4}{5}}% \tiny: 5/6
\or% 11pt
  \newcommand{\miniscule}{\@setfontsize\miniscule{5}{6}}% \tiny: 6/7
\or% 12pt
  \newcommand{\miniscule}{\@setfontsize\miniscule{5}{6}}% \tiny: 6/7
\fi
\makeatother

\DeclareRobustCommand{\optbar}[1]{\shortstack{{\miniscule (\rule[.5ex]{1.25em}{.18mm})}
  \\ [-.7ex] $#1$}}

\def\electron   {{\ensuremath{\Pe}}\xspace}
\def\en         {{\ensuremath{\Pe^-}}\xspace}   % electron negative (\em is taken)
\def\ep         {{\ensuremath{\Pe^+}}\xspace}
\def\epm        {{\ensuremath{\Pe^\pm}}\xspace}

\def\muon       {{\ensuremath{\Pmu}}\xspace}
\def\mup        {{\ensuremath{\Pmu^+}}\xspace}
\def\mun        {{\ensuremath{\Pmu^-}}\xspace} % muon negative (\mum is taken)

\def\tauon      {{\ensuremath{\Ptau}}\xspace}

\def\neub       {{\ensuremath{\overline{\Pnu}}}\xspace}
\def\W      {{\ensuremath{\PW}}\xspace}
\def\Wp     {{\ensuremath{\PW^+}}\xspace}
\def\Wm     {{\ensuremath{\PW^-}}\xspace}

\def\Z      {{\ensuremath{\PZ}}\xspace}

\def\cquark    {{\ensuremath{\Pc}}\xspace}

\def\bquark    {{\ensuremath{\Pb}}\xspace}

\def\tquark    {{\ensuremath{\Pt}}\xspace}
\def\tquarkbar {{\ensuremath{\overline \tquark}}\xspace}
\def\ttbar     {{\ensuremath{\tquark\tquarkbar}}\xspace}

  \def\Kbar    {{\kern 0.2em\overline{\kern -0.2em \PK}{}}\xspace}

\def\KorKbar    {\kern 0.18em\optbar{\kern -0.18em K}{}\xspace}

  \def\Dbar    {{\kern 0.2em\overline{\kern -0.2em \PD}{}}\xspace}

\def\DorDbar    {\kern 0.18em\optbar{\kern -0.18em D}{}\xspace}

\def\Bbar    {{\ensuremath{\kern 0.18em\overline{\kern -0.18em \PB}{}}}\xspace}

\def\BorBbar    {\kern 0.18em\optbar{\kern -0.18em B}{}\xspace}

  \def\Y#1S{\ensuremath{\PUpsilon{(#1S)}}\xspace}% no space before {...}!

\def\Lbar        {{\ensuremath{\kern 0.1em\overline{\kern -0.1em\PLambda}}}\xspace}
\def\LorLbar    {\kern 0.18em\optbar{\kern -0.18em \PLambda}{}\xspace}

\def\BF         {{\ensuremath{\cal B}}\xspace}

\def\to                 {\ensuremath{\rightarrow}\xspace}

\def\AT#1     {\ensuremath{A_{\mathrm{T}}^{#1}}\xspace}           % 2

\def\C#1      {\ensuremath{\mathcal{C}_{#1}}\xspace}                       % 9
\def\Cp#1     {\ensuremath{\mathcal{C}_{#1}^{'}}\xspace}                    % 7
\def\Ceff#1   {\ensuremath{\mathcal{C}_{#1}^{\mathrm{(eff)}}}\xspace}        % 9  
\def\Cpeff#1  {\ensuremath{\mathcal{C}_{#1}^{'\mathrm{(eff)}}}\xspace}       % 7
\def\Ope#1    {\ensuremath{\mathcal{O}_{#1}}\xspace}                       % 2
\def\Opep#1   {\ensuremath{\mathcal{O}_{#1}^{'}}\xspace}                    % 7

\newcommand{\tev}{\ifthenelse{\boolean{inbibliography}}{\ensuremath{~T\kern -0.05em eV}\xspace}{\ensuremath{\mathrm{\,Te\kern -0.1em V}}}\xspace}
\newcommand{\gev}{\ensuremath{\mathrm{\,Ge\kern -0.1em V}}\xspace}
\newcommand{\mev}{\ensuremath{\mathrm{\,Me\kern -0.1em V}}\xspace}
\newcommand{\kev}{\ensuremath{\mathrm{\,ke\kern -0.1em V}}\xspace}
\newcommand{\ev}{\ensuremath{\mathrm{\,e\kern -0.1em V}}\xspace}
\newcommand{\gevc}{\ensuremath{{\mathrm{\,Ge\kern -0.1em V\!/}c}}\xspace}
\newcommand{\mevc}{\ensuremath{{\mathrm{\,Me\kern -0.1em V\!/}c}}\xspace}
\newcommand{\gevcc}{\ensuremath{{\mathrm{\,Ge\kern -0.1em V\!/}c^2}}\xspace}
\newcommand{\gevgevcccc}{\ensuremath{{\mathrm{\,Ge\kern -0.1em V^2\!/}c^4}}\xspace}
\newcommand{\mevcc}{\ensuremath{{\mathrm{\,Me\kern -0.1em V\!/}c^2}}\xspace}

\def\mm   {\ensuremath{\rm \,mm}\xspace}

\def\mum  {\ensuremath{{\,\upmu\rm m}}\xspace}

\def\nb {\ensuremath{\rm \,nb}\xspace}

\def\pb {\ensuremath{\rm \,pb}\xspace}

\def\invfb   {\ensuremath{\mbox{\,fb}^{-1}}\xspace}

\newcommand{\chisqndf}{\ensuremath{\chi^2/\mathrm{ndf}}\xspace}

\def\gsim{{~\raise.15em\hbox{$>$}\kern-.85em
          \lower.35em\hbox{$\sim$}~}\xspace}
\def\lsim{{~\raise.15em\hbox{$<$}\kern-.85em
          \lower.35em\hbox{$\sim$}~}\xspace}

\def\ptot       {\mbox{$p$}\xspace}
\def\pt         {\mbox{$p_{\rm T}$}\xspace}
\def\pte        {\mbox{$p^e_{\rm T}$}\xspace}
\def\IT        {\mbox{$I^e_{\rm T}$}\xspace}
\def\ET        {\mbox{$E^{\gamma}_{\rm T}$}\xspace}
\def\ptch        {\mbox{$p^{\textup{ch}}_{\rm T}$}\xspace}
\def\etae        {\mbox{$\eta^e$}\xspace}

\def\fewz       {\mbox{\textsc{Fewz}}\xspace}

\def\geant      {\mbox{\textsc{Geant4}}\xspace}

\def\herwig     {\mbox{\textsc{Herwig++}}\xspace}

\def\pythia     {\mbox{\textsc{Pythia}}\xspace}
\def\dynnlo     {\mbox{\textsc{DYNNLO}}\xspace}
\def\resbos     {\mbox{\textsc{ResBos}}\xspace}

\def\tell1  {TELL1\xspace}
\def\ukl1   {UKL1\xspace}

%% file: title-LHCb-PAPER.tex
% $Id: title-LHCb-PAPER.tex 96357 2016-08-04 10:12:08Z sirendi $
% ===============================================================================
% Purpose: LHCb-PAPER journal paper title page template
% Author: 
% Created on: 2010-09-25
% ===============================================================================

%%%%%%%%%%%%%%%%%%%%%%%%%
%%%%%  TITLE PAGE  %%%%%%
%%%%%%%%%%%%%%%%%%%%%%%%%
\begin{titlepage}
\pagenumbering{roman}

% Header ---------------------------------------------------
\vspace*{-1.5cm}
\centerline{\large EUROPEAN ORGANIZATION FOR NUCLEAR RESEARCH (CERN)}
\vspace*{1.5cm}
\noindent
\begin{tabular*}{\linewidth}{lc@{\extracolsep{\fill}}r@{\extracolsep{0pt}}}
\ifthenelse{\boolean{pdflatex}}% Logo format choice
{\vspace*{-2.7cm}\mbox{\!\!\!\includegraphics[width=.14\textwidth]{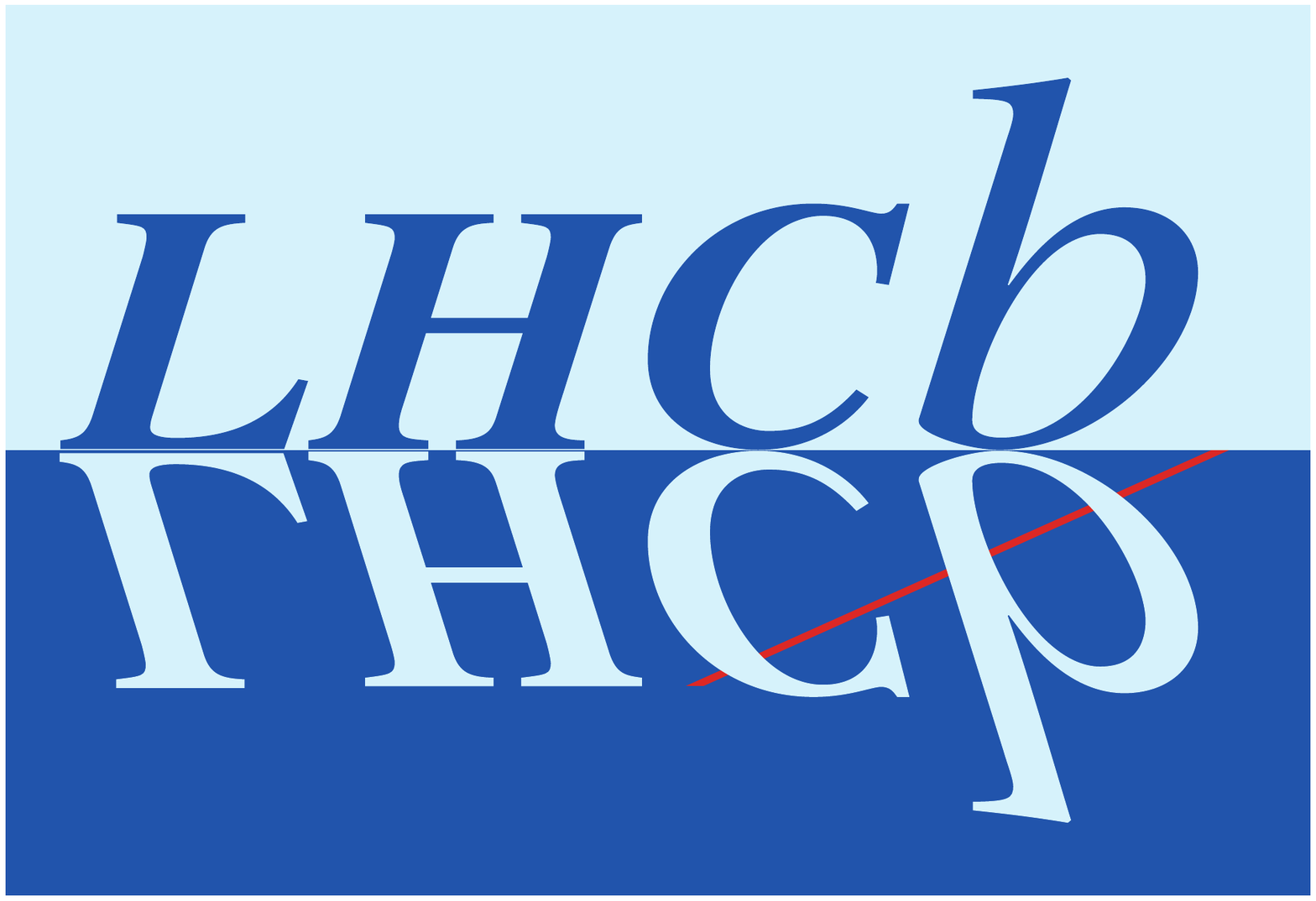}} & &}%
{\vspace*{-1.2cm}\mbox{\!\!\!\includegraphics[width=.12\textwidth]{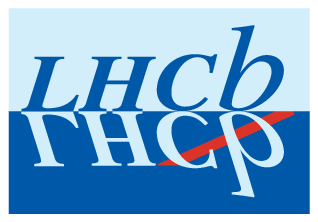}} & &}%
\\
 & & CERN-EP-2016-179 \\  % ID 
 & & LHCb-PAPER-2016-024 \\  % ID 
 & & October 20, 2016 \\
 & & \\
% not in paper \hline
\end{tabular*}

\vspace*{0.7cm}

% Title --------------------------------------------------
{\normalfont\bfseries\boldmath\huge
\begin{center}
  Measurement of forward $W\to\electron\nu$ production in $pp$ collisions \\ at $\sqrt{s}=8$\tev
\end{center}
}

\vspace*{0.7cm}

% Authors -------------------------------------------------
\begin{center}
%In the footnote, replace 'paper' by 'letter' in case of submission to PRL or PLB 
The LHCb collaboration\footnote{Authors are listed at the end of this paper.}
\end{center}

\vspace{\fill}

% Abstract -----------------------------------------------
\begin{abstract}
  \noindent
  A measurement of the cross-section for $\W\to\electron\nu$ production in $pp$ collisions is presented using data corresponding to an integrated luminosity of $2$\invfb collected by the LHCb experiment at a centre-of-mass energy of $\sqrt{s}=8$\tev. The electrons are required to have more than $20$\gev of transverse momentum and to lie between $2.00$ and $4.25$ in pseudorapidity. The inclusive \W production cross-sections, where the \W decays to $\electron\nu$, are measured to be
  \input{results/WxSec_chargedOnly.tex}
  \noindent where the first uncertainties are statistical, the second are systematic, the third are due to the knowledge of the LHC beam energy and the fourth are due to the luminosity determination.
  \noindent Differential cross-sections as a function of the electron pseudorapidity are measured. The \Wp/\Wm cross-section ratio and production charge asymmetry are also reported. Results are compared with theoretical predictions at next-to-next-to-leading order in perturbative quantum chromodynamics. Finally, in a precise test of lepton universality, the ratio of \W boson branching fractions is determined to be
	\input{results/leptonUniversality_limited.tex}
	\noindent where the first uncertainty is statistical and the second is systematic.
\end{abstract}

\vspace*{0.5cm}

\begin{center}
  Published in JHEP 10 (2016) 030 
\end{center}

\vspace{\fill}

{\footnotesize 
\centerline{\copyright~CERN on behalf of the \lhcb collaboration, licence \href{http://creativecommons.org/licenses/by/4.0/}{CC-BY-4.0}.}}
\vspace*{2mm}

\end{titlepage}

%%%%%%%%%%%%%%%%%%%%%%%%%%%%%%%%
%%%%%  EOD OF TITLE PAGE  %%%%%%
%%%%%%%%%%%%%%%%%%%%%%%%%%%%%%%%

%  empty page follows the title page ----
\newpage
\setcounter{page}{2}
\mbox{~}
%\newpage
%
%% Author List ----------------------------
%%  You need to get a new author list!
%\input{LHCb_authorlist.tex}
%
%The author list for journal publications is provided by the Membership Committee shortly after 'approval to go to paper' has been given.
%%It will be made available on the page
%%\verb!http://www.physik.uzh.ch/~strauman/forMemCo/LHCb-PAPER-XXXX-XXX/! .
%It will be sent to you by email shortly after a paper number has beens assigned.
%The author list should be included already at first circulation, 
%to allow new members of the collaboration to verify whether they have been included correctly.
%Occasionally a misspelled name is corrected or associated institutions become full members.
%In that case, a new author list will be sent to you.
%In case line numbering doesn't work well after including the authorlist, try moving the \verb!\bigskip! after the last author to a separate line.
%
%
%The authorship for Conference Reports should be ``The LHCb
%  collaboration'', with a footnote giving the name(s) of the contact
%  author(s), but without the full list of collaboration names.

\cleardoublepage

%% file: results/WxSec_chargedOnly.tex
\begin{align*}
\begin{split}
	\sigma_{\Wp\to\ep\nu_{\electron}}&=1124.4\pm 2.1\pm 21.5\pm 11.2\pm 13.0\pb,\\
	\sigma_{\Wm\to\electron^{-}\neub_{\electron}}&=\,\,\,809.0\pm 1.9\pm 18.1\pm\,\,\,7.0\pm \phantom{0}9.4\pb,
\end{split}
\end{align*}

%% file: results/leptonUniversality_limited.tex
\begin{align*}
\begin{split}
	\BF(\W\to\electron\nu)/\BF(\W\to\muon\nu)=1.020\pm 0.002\pm 0.019,
\end{split}
\end{align*}

%% file: introduction.tex
\section{Introduction}
\label{sec:introduction}

Precise measurements of the production cross-sections for \W and \Z bosons are important tests of the quantum chromodynamic (QCD) and electroweak (EW) sectors of the Standard Model (SM). In addition, the parton distribution functions (PDFs) of the proton can be better constrained~\cite{PDF4LHC}. The production of EW bosons has therefore been an important benchmark process to measure at current and past colliders. Measurements performed by the ATLAS~\cite{WZATLAS, ATLASDY8TeV, ATLAS13TeV}, CMS~\cite{CMSWZpT8TeV, CMSWasymmetry8TeV, CMSWZ8TeV}, and LHCb~\cite{WmunuLHCb, WZLHCb8TeV, ZmumuLHCb, ZeeLHCb, ZeeLHCb8TeV, ZtautauLHCb, Z13TeV} collaborations are in good agreement with theoretical predictions that are determined from parton-parton cross-sections convolved with PDFs. The precision of these predictions is limited by the accuracy of the PDFs and by unknown QCD corrections which are beyond next-to-next-to-leading order (NNLO) in perturbative QCD~\cite{FEWZ,FEWZ2}.

The PDFs, as functions of the Bjorken-$x$ values of the partons, have significant uncertainties at very low and large momentum fractions. Since the Bjorken-$x$ values of the interacting partons, $x_a$ and $x_b$, are related to the boson through its rapidity, \mbox{$y=\frac{1}{2}\ln{\frac{x_a}{x_b}}$}, forward measurements of production cross-sections are particularly valuable in constraining PDFs. The LHCb detector, which is instrumented in the forward region, is in a unique situation to provide input on determining accurate PDFs at small and large Bjorken-$x$ values. At large rapidities the measurements are mainly sensitive to scattering between valence and sea quarks, while at low rapidities scattering between pairs of sea quarks also contributes significantly. The \Wp/\Wm cross-section ratio and the production charge asymmetry of the \W boson are primarily sensitive to the ratio of $u$- and $d$-quark densities. In addition, the cross-section ratio and charge asymmetry enable the SM to be tested to greater precision since experimental and theoretical uncertainties partially cancel.

Here, the \W production cross-section is measured in the electron\footnote{When referred to generically, ``electron" denotes both \ep and \en.} final state. Compared to muons, the measurement of electrons has an additional experimental difficulty arising from the bremsstrahlung emitted when traversing the detector material. While the emitted photon energy can often be recovered for low-energy particles, electrons from \W boson decays tend to have high momentum, with bremsstrahlung photons that are not generally well-separated from the lepton. Coupled with the fact that individual LHCb calorimeter cells saturate by design at a transverse energy of approximately 10\gev, this leads to a poor energy measurement and a reconstructed distribution of transverse momentum, \pte, which differs significantly from the true transverse momentum of the electrons. In contrast, the electron direction is measured well, so that the differential cross-section in lepton pseudorapidity has negligible bin-to-bin migrations.

This paper presents measurements of the $\W\to\electron\nu$ cross-sections\footnote{The decay $\W\to\electron\nu$ denotes both $\Wp\to\ep\nu_e$ and $\Wm\to\electron^{-}\neub_{\electron}$ and similarly for the other leptonic decays. The $\W\to\electron\nu$ cross-section denotes the product of the cross-section for \W boson production and the branching fraction for $\W\to\electron\nu$ decay.}, cross-section ratios, and the charge asymmetry at $\sqrt{s}=8$\tev using data corresponding to an integrated luminosity of $2$\invfb collected by the LHCb detector. Measurements are made in eight bins of lepton pseudorapidity. The electrons are required to have more than $20$\gev of transverse momentum\footnote{Natural units with $\hbar=c=1$ are used throughout.} and to lie between $2.00$ and $4.25$ in pseudorapidity. The results are corrected for quantum electrodynamic (QED) final-state radiation (hereinafter denoted as ``Born level"). These requirements define the fiducial region of the measurements.

%% file: detector.tex
\section{Detector and simulation}
\label{sec:Detector}

The LHCb detector~\cite{Alves:2008zz,LHCb-DP-2014-002} is a single-arm forward spectrometer designed for the study of particles containing \bquark or \cquark quarks. The detector includes a high-precision tracking system consisting of a silicon-strip vertex detector surrounding the $pp$ interaction region, a large-area silicon-strip detector located upstream of a dipole magnet with a bending power of about $4{\mathrm{\,Tm}}$, and three stations of silicon-strip detectors and straw drift tubes placed downstream of the magnet. The tracking system provides a measurement of momentum, \ptot, of charged particles with a relative uncertainty that varies from 0.5\% at low momentum to 1.0\% at 200\gev. The minimum distance of a track to a primary vertex (PV), the impact parameter (IP), is measured with a resolution of $(15+29/\pt)\mum$, where \pt is the component of the momentum transverse to the beam, in\,\gev. Photons, electrons and hadrons are identified by a calorimeter system consisting of scintillating-pad (SPD) and preshower detectors (PRS), an electromagnetic calorimeter (ECAL) and a hadronic calorimeter (HCAL). The online event selection is performed by a trigger, which consists of a hardware stage, based on information from the calorimeter and muon systems, followed by a software stage, which applies a full event reconstruction. A set of global event cuts (GEC) is applied, which prevents events with high occupancy dominating the processing time of the software trigger.

Simulated data are used to optimise the event selection, estimate the background contamination and determine some efficiencies. In the simulation, $pp$ collisions are generated using \pythia 8~\cite{Sjostrand:2007gs,*Sjostrand:2006za} with a specific \lhcb configuration~\cite{LHCb-PROC-2010-056}. The interaction of the generated particles with the detector, and its response, are implemented using the \geant toolkit~\cite{Allison:2006ve, *Agostinelli:2002hh} as described in Ref.~\cite{LHCb-PROC-2011-006}. The momentum distribution of the partons inside the proton is parameterised by the leading-order \cteq~\cite{CTEQ6L1} PDF set. Final-state radiation (FSR) of the outgoing leptons is simulated using the model implemented internally within \pythia 8~\cite{pythiaFSR}.

%% file: candidateSelection.tex
\section{Event selection}
\label{sec:candidateSelection}

The production of $\W\to\electron\nu$ is characterised by a single, isolated high-\pt charged particle originating from a PV with a large energy deposit in the electromagnetic calorimeter. However, several other physics processes can mimic this experimental signature. Significant EW backgrounds include $\Z\to\electron\electron$ with one electron in the \lhcb acceptance,\footnote{\Z denotes the combined \Z and virtual photon ($\gamma^*$) contribution.} and $\Z\to\tau\tau$ and $\W\to\tau\nu$, where the $\tau$ decays to a final state containing an electron. Prompt photon production in association with jets contributes in cases where the photon converts to an $ee$ pair and only one electron is reconstructed and selected. Hadronic backgrounds stem from four sources: hadron misidentification (hereinafter denoted as ``fake electrons"), semileptonic heavy flavour decay, decay in flight, and \ttbar production.

The event selection requires the electron candidate to satisfy the trigger at both hardware and software levels. The reconstructed electron candidates should have pseudorapidity, \etae, between 2.00 and 4.25, have \pte in excess of 20\gev and should satisfy stringent track quality criteria. In particular, the relative uncertainty on the momentum is required to be less than 10\% to ensure that the charge is measured well. The upper limit of $\etae<4.25$ is imposed due to the limited acceptance of the calorimetry. To be identified as electrons, the candidates are required to deposit energy $E_{\textup{ECAL}}>0.15p^e$ in the ECAL while depositing relatively little energy $E_{\textup{HCAL}}<0.0075p^e$ in the HCAL, where $p^e$ is the momentum of the electron. The candidates are also required to have deposited energy of more than 50\mev in the PRS. The background formed by $\Z\to\electron\electron$ events with both electrons in the \lhcb acceptance is largely removed using a dedicated dielectron software trigger.

The remainder of the selection exploits other physical features of the process. Electrons from the \W boson decay are prompt, in contrast to leptons that come from decays of heavy flavour mesons or $\tau$ leptons. Hence the IP is required to be less than 0.04\mm. Another discriminant against hadronic processes is the fact that electrons from the \W boson tend to be isolated. On the other hand, leptons originating from hadronic decays, or fake electrons, tend to have hadrons travelling alongside them. The isolation requirement is set to be $\IT>0.9$, where \IT is defined as
\begin{equation}
\label{eq:isolation}
	\IT\equiv\frac{\pte}{\pte+\ET+\ptch}.
\end{equation}
\noindent Here \ET is the sum of the transverse component of neutral energy in the annular cone with \mbox{$0.1<R<0.5$}, where $R\equiv\sqrt{\Delta\eta^{2} + \Delta\phi^{2}}$ and $\Delta\eta$ and $\Delta\phi$ are the differences in the pseudorapidity and azimuthal angle between the candidate and the particle being considered, and \ptch is the scalar sum of the transverse momenta of charged tracks in the same annular cone. Bremsstrahlung photons are mostly contained in the range \mbox{$0.0<R<0.1$} and so are excluded from the isolation requirement.

%% file: eventYield.tex
\section{Signal yield}
\label{sec:eventYield}

In total, $1\,368\,539$ $\W\to\electron\nu$ candidates fulfil the selection requirements. The signal yields are determined in eight bins of lepton pseudorapidity and for each charge. Binned maximum likelihood template fits to the \pt distribution of the electron candidate are performed in the range $20<\pte<65\gev$, following Ref.~\cite{fitting}. The \pte spectra in the 16 bins of pseudorapidity and charge with the results of the fits superimposed are reported in Appendix~\ref{sec:leptonptFits}.

Templates for $\W\to\electron\nu$, $\W\to\tau\nu$, $\Z\to\electron\electron$ and $\Z\to\tau\tau\to\electron X$ are taken from simulation, where $X$ represents any additional particles. The known ratio of branching fractions~\cite{PDG} is used to constrain the ratio of $\W\to\tau\nu$ to $\W\to\electron\nu$. The measured \lhcb cross-section for $\Z\to\mu\mu$ production~\cite{WZLHCb8TeV} is used to constrain $\Z\to\electron\electron$ and $\Z\to\tau\tau\to\electron X$ in the fit, and knowledge of the ratio of branching fractions to different leptonic final states of the \Z boson~\cite{PDG} is also taken into account.

Contributions from $\W\gamma$, $\Z\gamma$, $\W\W$, $\W\Z$, and $\ttbar$ events are included in the fits. These processes account for $(0.46\pm0.01)\%$ of the selected candidates and are denoted as ``rare processes" in the following. The templates for these processes are obtained from simulation and normalised to the \mcfm~\cite{MCFM} NLO cross-section predictions.

The production of prompt photons in association with jets has a cross-section of about 50\nb for a $\pt>20\gev$ photon within the \lhcb acceptance, as computed using \mcfm at NLO. This process mimics the signal in cases where the photon converts into an $ee$ pair in the detector material and one electron satisfies the $\W\to\electron\nu$ selection. A sample of photon+jets candidates is obtained from data by searching for an $ee$ pair with mass below 50\mev and applying stringent selection criteria to the candidates. Simulation is used to account for the differences in the $\W\to\electron\nu$ and $\gamma\to ee$ selections.

Hadron misidentification occurs when hadrons begin to shower early in the ECAL, giving a shower profile similar to that of electrons. These hadrons, however, will tend to deposit fractionally more energy in the HCAL than genuine electrons and will also be less isolated on average. A template for the \pt distribution of fake electrons is determined using data, by modifying the isolation and HCAL energy requirements of the selection to produce a sample dominated by hadrons.

The semileptonic decay of heavy flavour (HF) hadrons gives rise to genuine electrons. This background is suppressed using the IP requirement to exploit the long lifetimes of hadrons containing \bquark and \cquark quarks. The remaining HF component is described by a data-driven template obtained by applying the standard selection but requiring the impact parameter to be significantly different from zero. The normalisation of the remaining contribution in the fit to \pte is determined from a separate template fit to the $\chi^2_{\textup{IP}}$ distribution, where $\chi^2_{\textup{IP}}$ is the difference between the $\chi^2$ of the PV fit when reconstructed with and without the candidate electron. The fractional HF component in the signal region is determined to be smaller than 0.8\% at 68\% confidence level.

\begin{figure}[tb]
\begin{center}
	\includegraphics[width=0.8\linewidth]{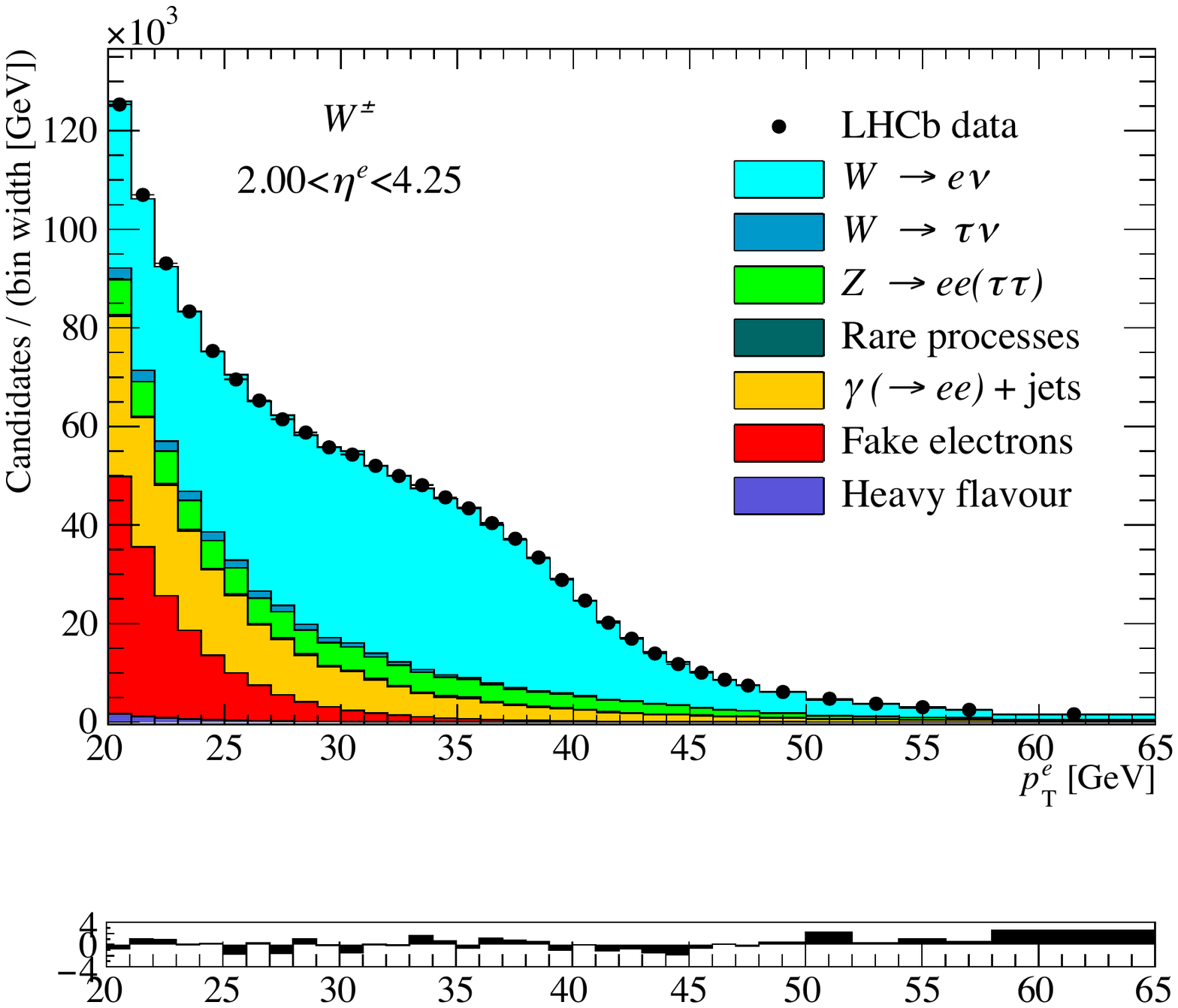}
\vspace*{-0.5cm}
\end{center}
\caption{
	The inclusive fit to the \pte distribution of the full dataset. The \chisqndf of the fit is 1.1 with 33 degrees of freedom.}
\label{fig:incFit}
\end{figure}
The $\W\to(\electron,\tauon)\nu_{(\electron,\tau)}$ and fake electron fractions are free to vary in the fits, while the remaining components are constrained as described previously. The validity of the SM is implicitly assumed in the constraints based on theoretical cross-sections obtained from \mcfm and in extracting template shapes from simulation. The $\Wp\to\ep\nu_e$ and $\Wm\to\electron^{-}\neub_{\electron}$ sample purities are determined to be $(63.95\pm0.19)\%$ and $(56.06\pm0.21)\%$. The \pte distribution of the full dataset with the result of the fit overlaid is shown for illustration in Fig.~\ref{fig:incFit} and is used in the estimation of systematic uncertainties.

%% file: crosssectionMeasurement.tex
\section{Cross-section measurement}
\label{sec:crosssectionMeasurement}

The production cross-section for $\W\to\electron\nu$ is measured in each bin of lepton pseudorapidity and for each charge with electron transverse momentum in excess of 20\gev. The cross-section is determined from

\begin{equation}
	\sigma_{i}^{\W\to\electron\nu}=\frac{N_{i}^{\W}}{\mathcal{A}_{i}\,L\,\epsilon_{i}^{\textup{tot}}\,f^{\textup{FSR}}},
\end{equation}
\noindent where $N^{\W}_{i}$ is the signal yield in the range $20<\pte<65\gev$ obtained from the fit in bin $i$ of $\eta^{\electron}$, $\epsilon_{i}^{\textup{tot}}$ is the total efficiency in that bin, and $L$ is the integrated luminosity. The signal yields are corrected for excluded candidates with $\pte>65\gev$ by computing a charge-dependent acceptance factor, $\mathcal{A}_{i}$, using a \resbos~\cite{RESBOS1,RESBOS2,RESBOS3} simulation.

The results of the measurement are quoted at Born level to enable comparisons to theoretical predictions that do not incorporate the effect of QED final-state radiation. Correcting to Born level also enables a comparison to be made with the measurement of $\W\to\mu\nu$. Corrections due to FSR, $f^{\textup{FSR}}$, are computed separately using \pythia 8 and \herwig~\cite{herwig} and then averaged. The corrections are listed in Appendix~\ref{sec:tabulatedResults} so that the measurement can be compared to a prediction that incorporates the effect of FSR.

The total efficiency used to correct the candidate yield can be written as the product

\begin{equation}
	\epsilon^{\textup{tot}}\equiv\epsilon^{\textup{track}}\cdot\epsilon^{\textup{kin}}\cdot\epsilon^{\textup{PID}}\cdot\epsilon^{\textup{GEC}}\cdot\epsilon^{\textup{trigger}}\cdot\epsilon^{\textup{tight}}.
\end{equation}
\noindent The description and estimation of the various terms are explained below. Each subsequent efficiency is determined in a subset of events defined by the preceding requirements in order to ensure that correlations between the requirements are correctly accounted for.

The track reconstruction efficiency, $\epsilon^{\textup{track}}$, is the probability that an electron is reconstructed as a track satisfying standard track quality criteria and the requirement that the relative momentum uncertainty is less than 10\%. The efficiency is determined using simulation of $\W\to\electron\nu$ and cross-checked with a data-driven study using $\Z\to\electron\electron$ candidate events~\cite{ZeeLHCb8TeV}.

An electron with true \pt of more than 20\gev can be reconstructed as having \mbox{$\pte<20\gev$}. This is predominantly due to bremsstrahlung. For high-\pt candidates, the photons tend to lie close to the electron and are often not correctly identified by bremsstrahlung recovery. The correction for this effect, $\epsilon^{\textup{kin}}$, is determined using simulation and is cross-checked in data using the method outlined in Ref.~\cite{ZeeLHCb8TeV}.

Simulation of $\W\to\electron\nu$ is used to extract an efficiency, $\epsilon^{\textup{PID}}$, for the loose particle identification (PID) requirements that are applied in the initial selection of electron candidates. The efficiency is corrected using the data-driven technique employed for $\Z\to\electron\electron$ candidate events~\cite{ZeeLHCb8TeV}.

The hardware trigger incorporates a global event cut (GEC) on the number of SPD hits, $N_{\textup{SPD}}<600$, to prevent high-multiplicity events from dominating the processing time at trigger-level. Dimuon events have a less stringent requirement of $N_{\textup{SPD}}<900$ and are used to determine the fraction of events, $\epsilon^{\textup{GEC}}$, below $N_{\textup{SPD}}=600$. However, dimuon candidate events are not entirely comparable to $\W\to\electron\nu$ as electrons will shower in the detector and lead to more hits in the SPD. Nevertheless, after a suitable shift of the dimuon distribution, good agreement is observed with $\W\to\electron\nu$ candidate events.

A tag-and-probe method~\cite{ZeeLHCb8TeV} is used on $\Z\to\electron\electron$ data to determine the efficiency, $\epsilon^{\textup{trigger}}$, for the single-electron triggers. The tag is an electron from a \Z candidate that satisfies the above requirements and meets all trigger requirements. The probe is then used to determine the fraction of candidates that satisfy the trigger requirements. The hadronic background in the $\Z\to\electron\electron$ dataset is estimated using same-sign, $\epm\epm$, events. The efficiency for a veto on the dielectron trigger is determined using simulation of $\W\to\electron\nu$ and is close to 100\%.

Tight selection requirements consist of more stringent track quality requirements and PID requirements, as well as ensuring the track is prompt and isolated. The efficiency for these requirements, $\epsilon^{\textup{tight}}$, is determined using \Z data analogously to the procedure for determining the trigger efficiency.

Efficiencies determined from $\Z\to\electron\electron$ cannot be directly used for \W production due to the different couplings at the production and decay vertices, a different mixture of interacting quarks, and, most importantly, the difference in mass. This results in a \pte distribution that is harder for electrons from the \Z boson. Consequently, efficiencies that show a dependence on \pte are liable to be biased. This is corrected for in each bin of \etae using \W and \Z simulation.

%% file: systematicUncertainties.tex
\section{Systematic uncertainties}
\label{sec:systematicUncertainties}

\input{results/Table1.tex}
Several sources of systematic uncertainty affect the measurement. These are summarised in Table~\ref{tab:xSecSystematics} for the total cross-sections in the fiducial region and the ratio measurements where $R_{W^{\pm}}\equiv\sigma_{\Wp\to\ep\nu_{\electron}}/\sigma_{\Wm\to\electron^{-}\neub_{\electron}}$.

The yields determined from fits to the \pte distribution are affected by two types of uncertainty. The effect of the statistical uncertainty in the templates is evaluated using pseudoexperiments and is denoted as ``Yield (statistical)" in Table~\ref{tab:xSecSystematics}. All other sources of uncertainty in the fits are considered systematic in nature (denoted as ``Yield (systematic)" in Table~\ref{tab:xSecSystematics}) and are described in the next paragraph.

Templates for contributions from photon+jets, fake electrons and heavy flavours, determined using data, contain a mixture of physical processes. A simulation-based estimate for EW contamination is subtracted and a 50\% systematic uncertainty is assigned for the procedure. Components that are constrained in the fits are varied according to their respective uncertainties. Templates for $\Z\to\electron\electron$ and $\Z\to\tau\tau\to\electron X$ are subject to an uncertainty on the cross-section, and the normalisation of the rare processes has an uncertainty from the cross-sections and the luminosity determination. Two alternative control regions are considered for determining the fake electron component resulting in an uncertainty of 0.6\% on the total cross-section. The fits are repeated with these alternative regions to ascertain the uncertainty associated with the fake electron template. The systematic uncertainty on the normalisation of the heavy flavour component is 0.8\% and the data-driven \pt template is varied accordingly. The transverse momentum of the candidate in simulation is sensitive to both the potential mismodelling of track reconstruction and the description of the material traversed by the candidate. The latter affects the number of bremsstrahlung photons emitted and thus has an impact on the \pte of the candidate and, by extension, on the fits. Any potential mismodelling can be described by a scaling of the momentum, as explained in Ref.~\cite{ZeeLHCb8TeV}. The effect of varying the momentum scale on all simulation-based templates is tested on the inclusive fit shown in Fig.~\ref{fig:incFit} and the best fit value for the momentum scale is seen to be consistent with unity, suggesting that material in the detector is modelled well. An uncertainty of 0.5\% assigned on the momentum scale in Ref.~\cite{ZeeLHCb8TeV} is found to be appropriate for the measurement. Varying the momentum scale by its uncertainty in the fits binned in \etae leads to an uncertainty of 1.3\% on the total cross-section which is the largest contribution to ``Yield (systematic)".

The statistical uncertainty on the total efficiency is taken as a contribution to the uncertainty on the measurement and is denoted as ``Efficiency (statistical)" in Table~\ref{tab:xSecSystematics}. In the case of cross-sections, the uncertainties from the finite statistics of the \Z data and \Z/\W simulated samples all contribute. For the determination of the cross-section ratio and the charge asymmetry, only the uncertainty due to the simulation of the \W must be accounted for. All other sources of uncertainty in the efficiencies are collectively denoted as ``Efficiency (systematic)" in Table~\ref{tab:xSecSystematics} and are described in the next paragraph.

Data-driven cross-checks performed on the efficiencies determined using simulation lead to an uncertainty of 0.5\% on the track reconstruction efficiency, an uncertainty of 0.6\% on the kinematic efficiency due to the modelling of bremsstrahlung in simulation, and an uncertainty of 0.6\% on PID requirements. The statistical component of the uncertainty on the GEC efficiency is found to be 0.09\%. Since $\epsilon^{\textup{GEC}}$ is dependent on the number of primary vertices, $N_{\textup{PV}}$, the efficiency is measured separately for $N_{\textup{PV}}=\{1, 2, 3, \geq 4\}$ and combined. This is compared with the estimate of the efficiency obtained inclusively for all numbers of primary vertices and an uncertainty of 0.33\% is assigned based on the difference between the two methods. Overall, a systematic uncertainty of 0.34\% is assigned for the procedure to determine the efficiency from dimuon candidate events. An additional systematic uncertainty is assigned on the cross-sections, the cross-section ratio, and the charge asymmetry to account for the differences observed between electrons and positrons in simulation. Same-sign subtraction is performed when the $\Z\to\electron\electron$ data sample is used. A study that formed electron and charged pion combinations and counted opposite- and same-sign pairs~\cite{ZeeLHCb8TeV} leads to a systematic uncertainty of 0.25\% on the $\W\to\electron\nu$ cross-section due to the normalisation of hadronic contamination in the sample of $\Z\to\electron\electron$ candidates.

Half the difference between \pythia 8 and \herwig predictions is taken as the systematic component of the uncertainty on FSR corrections.

The statistical uncertainty on the acceptance corrections arises from the \resbos $\W$ simulated sample. Half the difference between \pythia 8 and \resbos is taken as a systematic uncertainty on a bin-by-bin basis and is assumed to be correlated between bins.

A small fraction of candidate electrons have the wrong charge assigned to them, which leads to a bias in the cross-section ratio and the charge asymmetry. A correction of $(0.58\pm0.05)\%$ is determined using simulation and applied to the measurements.

The uncertainty on the LHC beam energy at 8\tev~\cite{beam} leads to a relative uncertainty on the $\Wp\,(\Wm)$ cross-section of $1.00\,(0.86)\%$ determined using \dynnlo~\cite{dynnlo}. The uncertainty on the luminosity is 1.16\% for the 8\tev dataset~\cite{luminosity}.

%% file: results/Table1.tex
\begin{table}[t]
\begin{center}
\begin{tabular}{lccc}
Source & \multicolumn{3}{c}{Uncertainty [\%]} \\
\hline
&$\sigma_{\Wp\to\ep\nu_{\electron}}$&$\sigma_{\Wm\to\electron^{-}\neub_{\electron}}$&$R_{W^{\pm}}$ \\
\hline
Statistical$^{\dagger}$&0.19&0.24&0.30\\
\hline
Yield (statistical)$^{\dagger}$&0.28&0.40&0.48\\
Yield (systematic)&1.42&1.79&0.51\\
Efficiency (statistical)$^{\dagger}$&0.55&0.55&0.21\\
Efficiency (systematic)&1.11&1.14&0.54\\
FSR corrections$^{\dagger}$&0.05&0.07&0.09\\
Acceptance corrections (statistical)$^{\dagger}$&0.00&0.01&0.01\\
Acceptance corrections (systematic)&0.15&0.15&0.00\\
Charge mis-identification$^{\dagger}$&---&---&0.02\\
\hline
Systematic&1.91&2.23&0.91\\
\hline
Beam energy&1.00&0.86&0.14\\
Luminosity&1.16&1.16&---\\
\hline
Total&2.46&2.67&0.97\\
\end{tabular}
\end{center}
\caption{Summary of the relative uncertainties on the $\Wp$ and $\Wm$ boson cross-sections and on the cross-section ratio. Uncertainties marked with $^{\dagger}$ are assumed to be uncorrelated between bins; all others are taken to be correlated.}
\label{tab:xSecSystematics}
\end{table}

%% file: results.tex
\section{Results}
\label{sec:results}

\subsection{Propagation of uncertainties}
\label{subsec:uncertaintyPropagation}

When computing derived quantities such as the total cross-section, cross-section ratios, and the charge asymmetry, correlations between the 16 measurements of $\W\to\electron\nu$ in bins of \etae must be accounted for. Uncertainties marked with $^{\dagger}$ in Table~\ref{tab:xSecSystematics} are statistical in nature and are assumed to be uncorrelated between charges and bins of \etae. All other sources of systematic uncertainty are varied by one standard deviation around their nominal value for each measurement and the correlation between each pair of measurements is computed. Correlation matrices between bins of \etae for \Wp, \Wm, and \Wp against \Wm are reported in Appendix~\ref{sec:correlationCoefficients}. A consequence of the sizeable positive correlations is that many of the systematic uncertainties add coherently when integrating over bins, but partially cancel in determining \Wp/\Wm ratios.

Sect.~\ref{subsec:leptonUniversality} presents the ratio of the $\W\to\electron\nu$ and $\W\to\muon\nu$ branching fractions. Here, the systematic uncertainties of the respective measurements are taken to be uncorrelated between the two final states apart from the uncertainties on the GEC efficiency and the acceptance correction, which are taken to be fully correlated.

\subsection{Inclusive results}
\label{subsec:inclusiveResults}

Total inclusive cross-sections for $W\to\electron\nu$ production are obtained by summing the cross-sections in bins of \etae. The Born level cross-sections in the fiducial region defined as \mbox{$2.0<\eta^{\electron}<4.25$} and more than 20\gev of transverse momentum are measured to be
\input{results/WxSec.tex}
\noindent where the first uncertainties are statistical, the second are systematic, the third are due to the knowledge of the LHC beam energy and the fourth are due to the luminosity determination.

The \Wp to \Wm cross-section ratio is determined to be
\input{results/ratio.tex}
\noindent where uncertainties are statistical, systematic and due to the LHC beam energy measurement, respectively.

\subsection{Cross-sections as a function of electron pseudorapidity}
\label{subsec:differentialResults}

\begin{figure}[tb]
\begin{center}
	\includegraphics[width=0.78\linewidth]{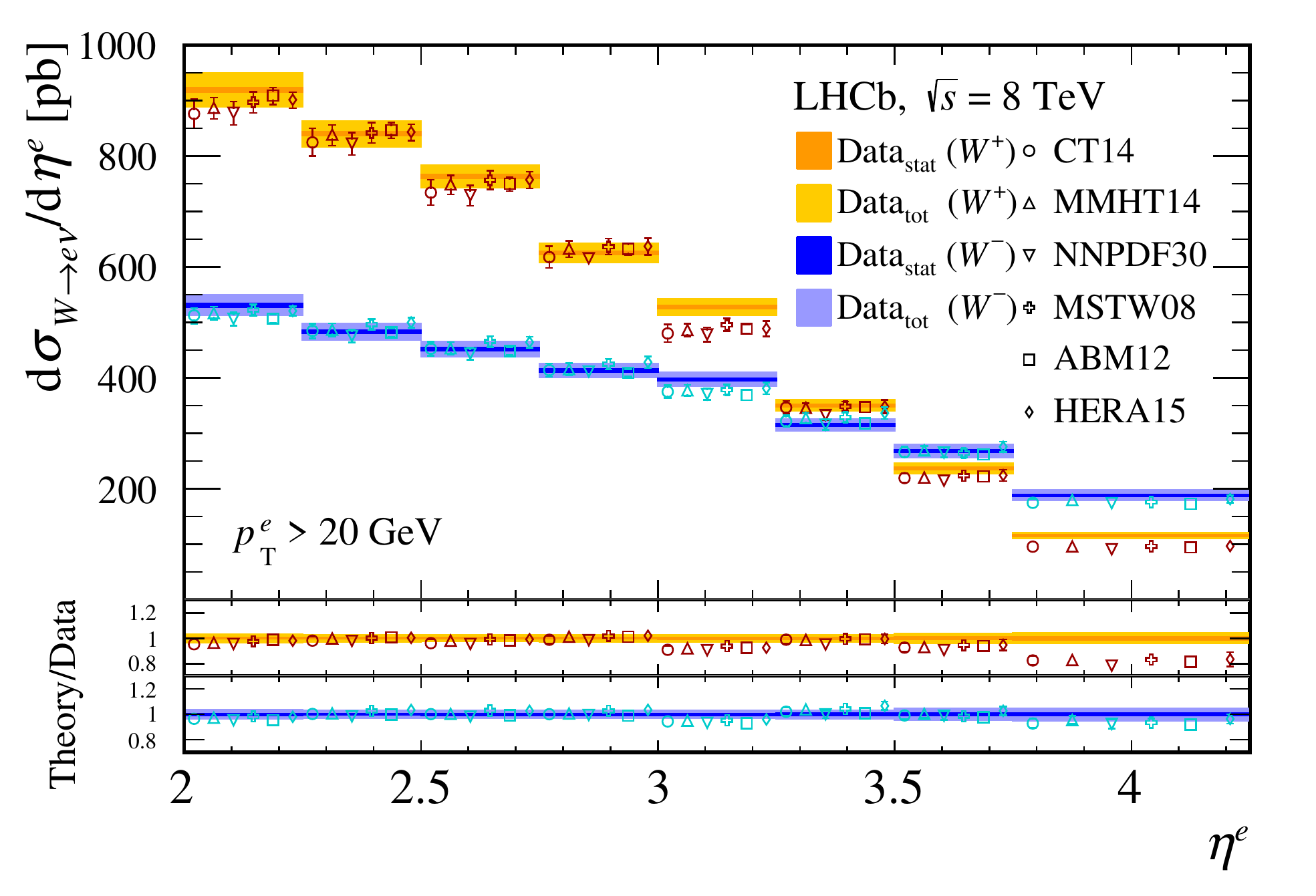}
\vspace*{-0.5cm}
\end{center}
\caption{
	The differential \Wp and \Wm cross-sections in bins of $\eta^{\electron}$. Measurements, represented as bands, are compared to NNLO predictions with different parameterisations of the PDFs (markers are displaced horizontally for presentation). The bottom panel displays the theory predictions divided by the measured cross-sections.}
\label{fig:diffXsec}
\end{figure}
Born level cross-sections as a function of electron pseudorapidity are tabulated in Appendix~\ref{sec:tabulatedResults}. The differential cross-sections as a function of $\eta^{\electron}$ are also determined and shown in Fig.~\ref{fig:diffXsec}. Measurements are compared to theoretical predictions calculated with the \fewz~\cite{FEWZ,FEWZ2} generator at NNLO for the six PDF sets: ABM12~\cite{ABM}, CT14~\cite{CT14}, HERA1.5~\cite{HERAPDF}, MMHT14~\cite{MMHT}, MSTW08~\cite{MSTW}, and NNPDF3.0~\cite{NNPDF}. Satisfactory agreement is observed apart from in the far forward region of the \Wp differential measurement, where the PDF uncertainties are also greatest.
\clearpage
\subsection{Cross-section ratio and charge asymmetry}
\label{subsec:RWandChargeAsymmetry}

\begin{figure}[tb]
\begin{center}
	\includegraphics[width=0.78\linewidth]{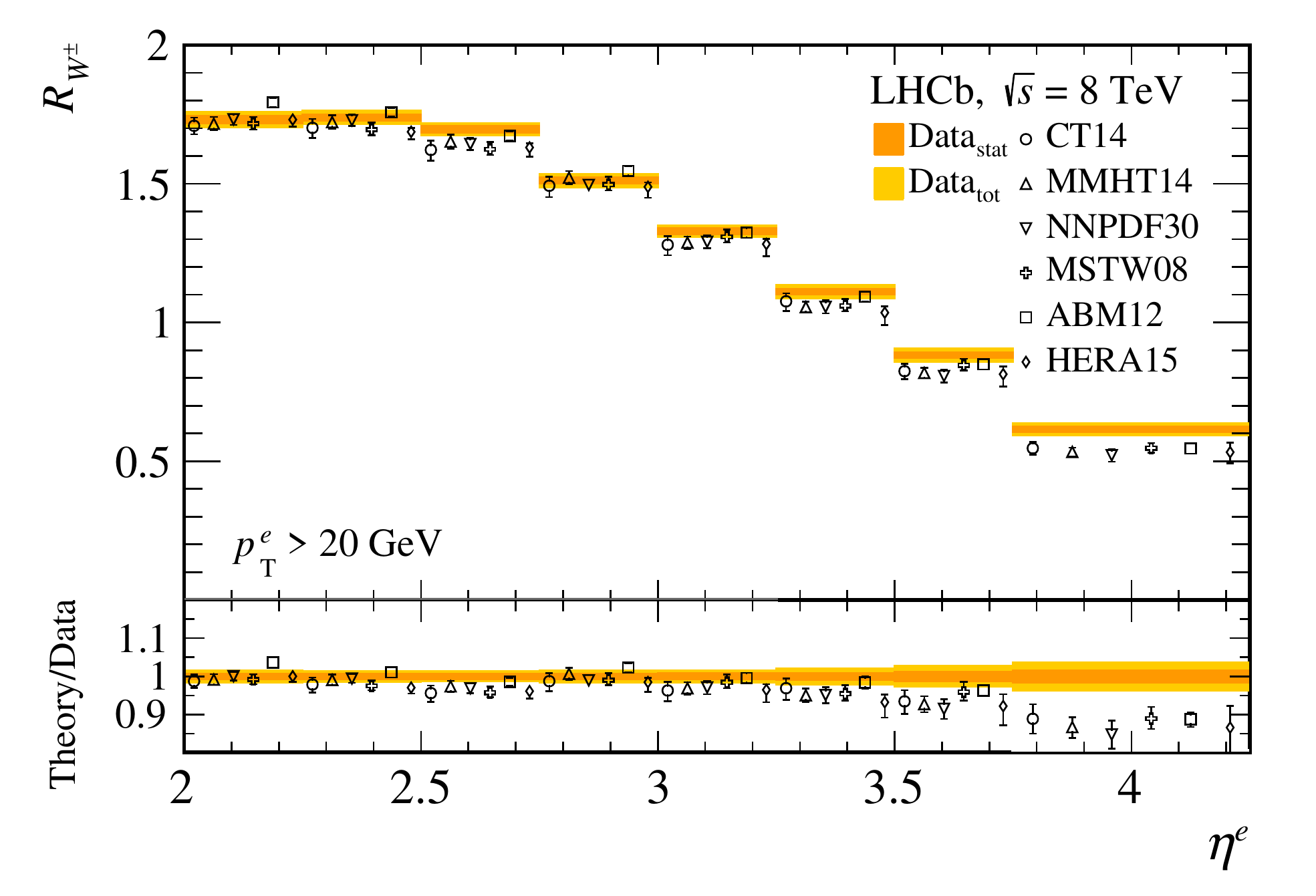}
\vspace*{-0.5cm}
\end{center}
\caption{
	The \Wp to \Wm cross-section ratio in bins of $\eta^{\electron}$. Measurements, represented as bands, are compared to NNLO predictions with different parameterisations of the PDFs (markers are displaced horizontally for presentation). The bottom panel displays the theory predictions divided by the measured cross-section ratios.}
\label{fig:xSecRatio}
\end{figure}
Cross-section ratios as a function of \etae are compared to theory predictions in Fig.~\ref{fig:xSecRatio} and the measurements are tabulated in Appendix~\ref{sec:tabulatedResults}. Overall the measurements are in agreement with theory predictions, with the exception of the far forward region. In this region the measured ratio is higher than the expectation as a consequence of the discrepancy seen in the \Wp cross-section in that region.

\begin{figure}[tb]
\begin{center}
	\includegraphics[width=0.78\linewidth]{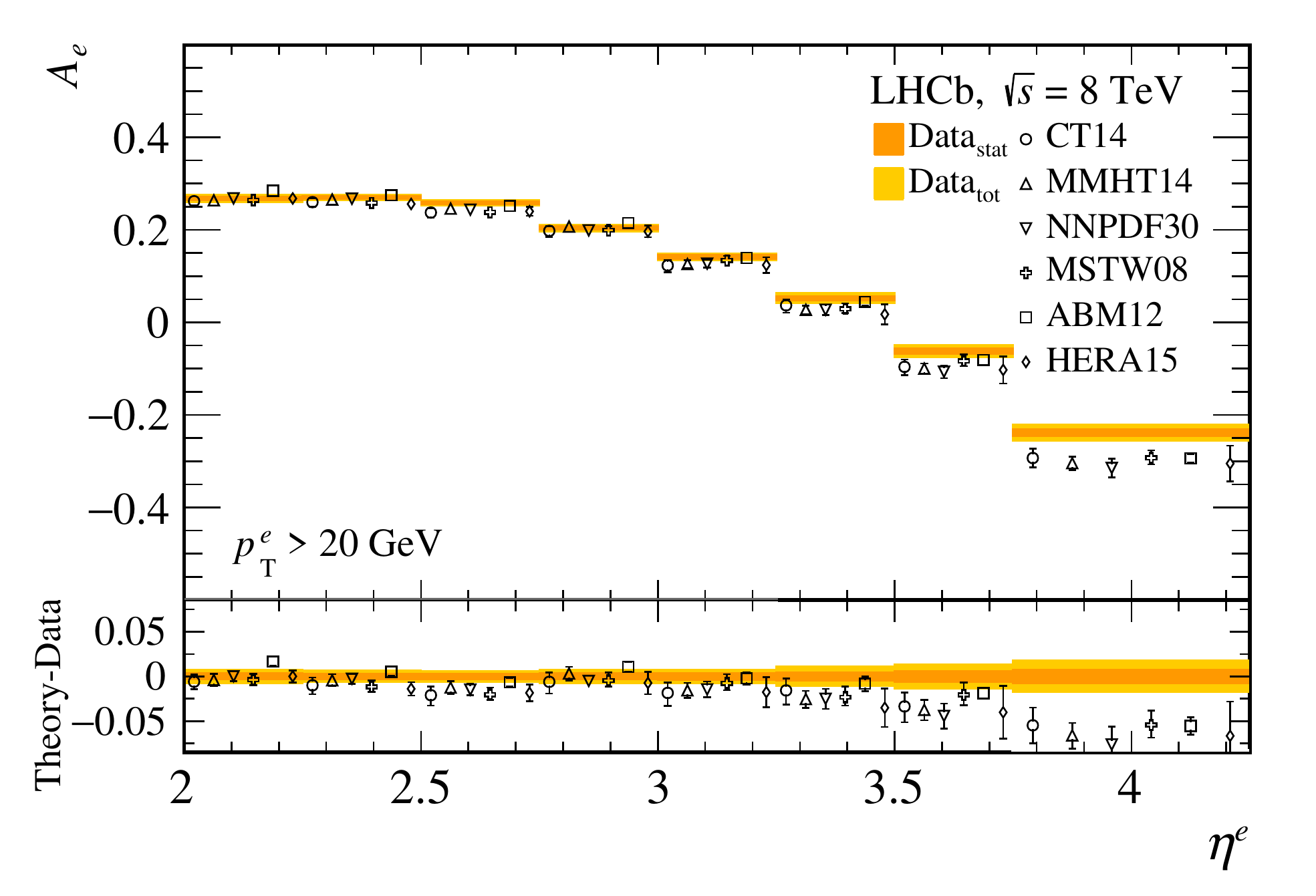}
\vspace*{-0.5cm}
\end{center}
\caption{
	The \W boson production charge asymmetry in bins of $\eta^{\electron}$. Measurements, represented as bands, are compared to NNLO predictions with different parameterisations of the PDFs (markers are displaced horizontally for presentation). The bottom panel displays the difference between theory predictions and the measured charge asymmetry.}
\label{fig:asymmetry}
\end{figure}
The \W boson production charge asymmetry is defined as

\begin{align}
	\begin{split}
		A_{\electron}\equiv\frac{\sigma_{\Wp\to\ep\nu_{\electron}}-\sigma_{\Wm\to\electron^{-}\neub_{\electron}}}{\sigma_{\Wp\to\ep\nu_{\electron}}+\sigma_{\Wm\to\electron^{-}\neub_{\electron}}}. \\ 
	\end{split}
\end{align}

\noindent The asymmetry is compared to theory predictions in bins of \etae in Fig.~\ref{fig:asymmetry}. The measurements are tabulated in Appendix~\ref{sec:tabulatedResults}.

\subsection{Lepton universality}
\label{subsec:leptonUniversality}

\begin{figure}[tb]
\begin{center}
	\includegraphics[width=0.78\linewidth]{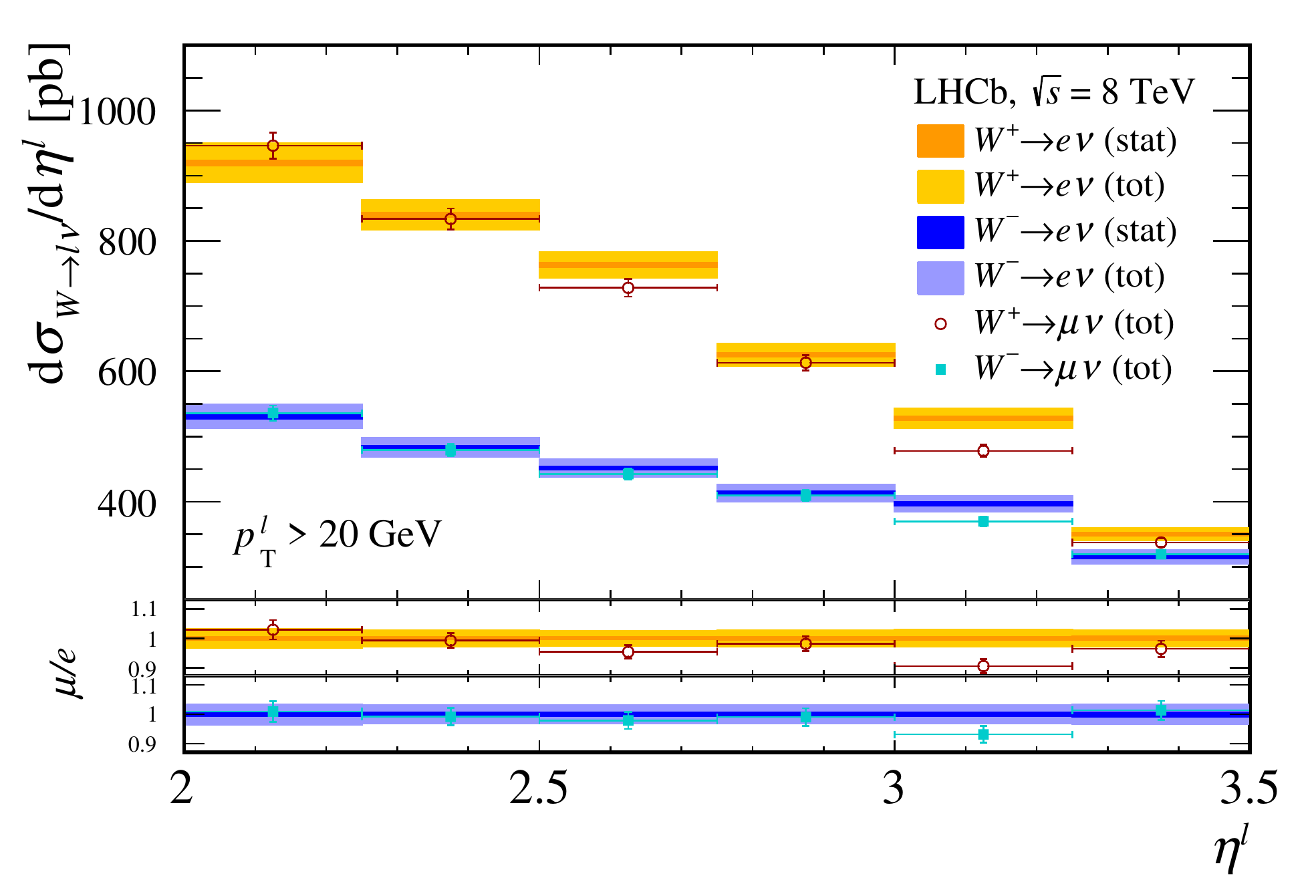}
\vspace*{-0.5cm}
\end{center}
\caption{
	The differential \Wp and \Wm cross-sections in bins of $\eta^{l}$. The measurement using electrons, represented as bands, is compared to the measurement in the muon final state. The bottom panel displays the muon results divided by the measurements in the electron final state.}
\label{fig:diffXsecmuComp}
\end{figure}
\begin{figure}[tb]
\begin{center}
	\includegraphics[width=0.78\linewidth]{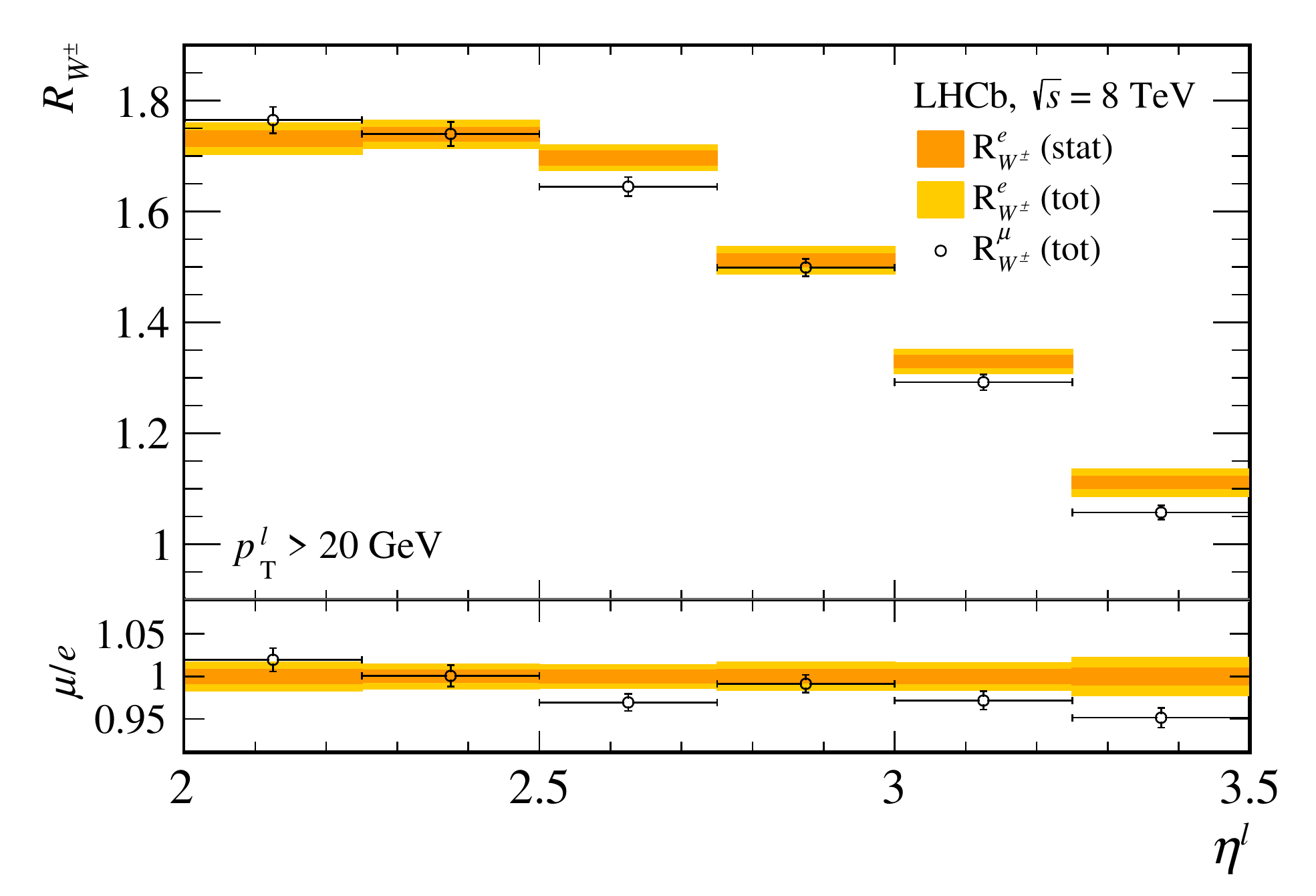}
\vspace*{-0.5cm}
\end{center}
\caption{
	The \Wp to \Wm cross-section ratio in bins of $\eta^{l}$. The measurement using electrons, represented as bands, is compared to the measurement in the muon final state. The bottom panel displays the muon results divided by the measurements in the electron final state.}
\label{fig:xSecRatiomuComp}
\end{figure}
\begin{figure}[tb]
\begin{center}
	\includegraphics[width=0.78\linewidth]{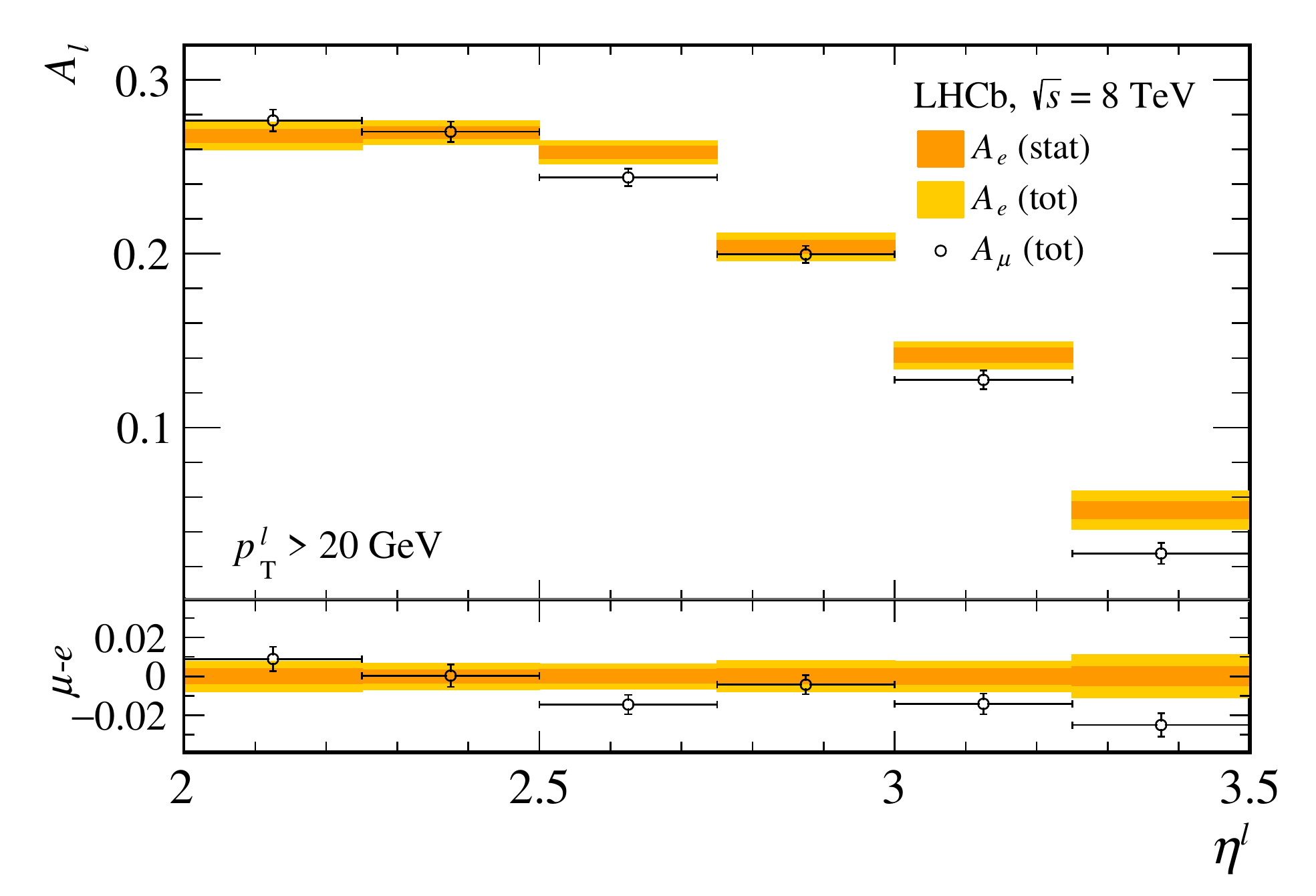}
\vspace*{-0.5cm}
\end{center}
\caption{
	The \W boson production charge asymmetry in bins of $\eta^{l}$. The measurement using electrons, represented as bands, is compared to the measurement in the muon final state. The bottom panel displays the difference between the muon and electron final states.}
\label{fig:asymmetrymuComp}
\end{figure}

Production of \W bosons in the forward region has also been studied in the muon final state~\cite{WZLHCb8TeV}. The muon measurement had a different upper kinematic limit in pseudorapidity, and consequently the bin boundaries only coincide with the present measurement for $\eta^{l}<3.50$. The results are therefore compared in the range $2.00<\eta^{l}<3.50$ as is shown in Figs.~\ref{fig:diffXsecmuComp}, \ref{fig:xSecRatiomuComp}, and \ref{fig:asymmetrymuComp}. The results of these measurements are seen to be consistent with the $\W\to\mu\nu$ measurements and no significant deviation from lepton universality is observed once uncertainties and correlations between measurements are taken into account. Fig.~\ref{fig:diffXsecmuComp} shows good agreement, apart from the bin $3.00<\eta^{l}<3.25$ for \Wp, where the difference is approximately 3 standard deviations.
\clearpage
\begin{figure}[tb]
\begin{center}
	\includegraphics[width=0.8\linewidth]{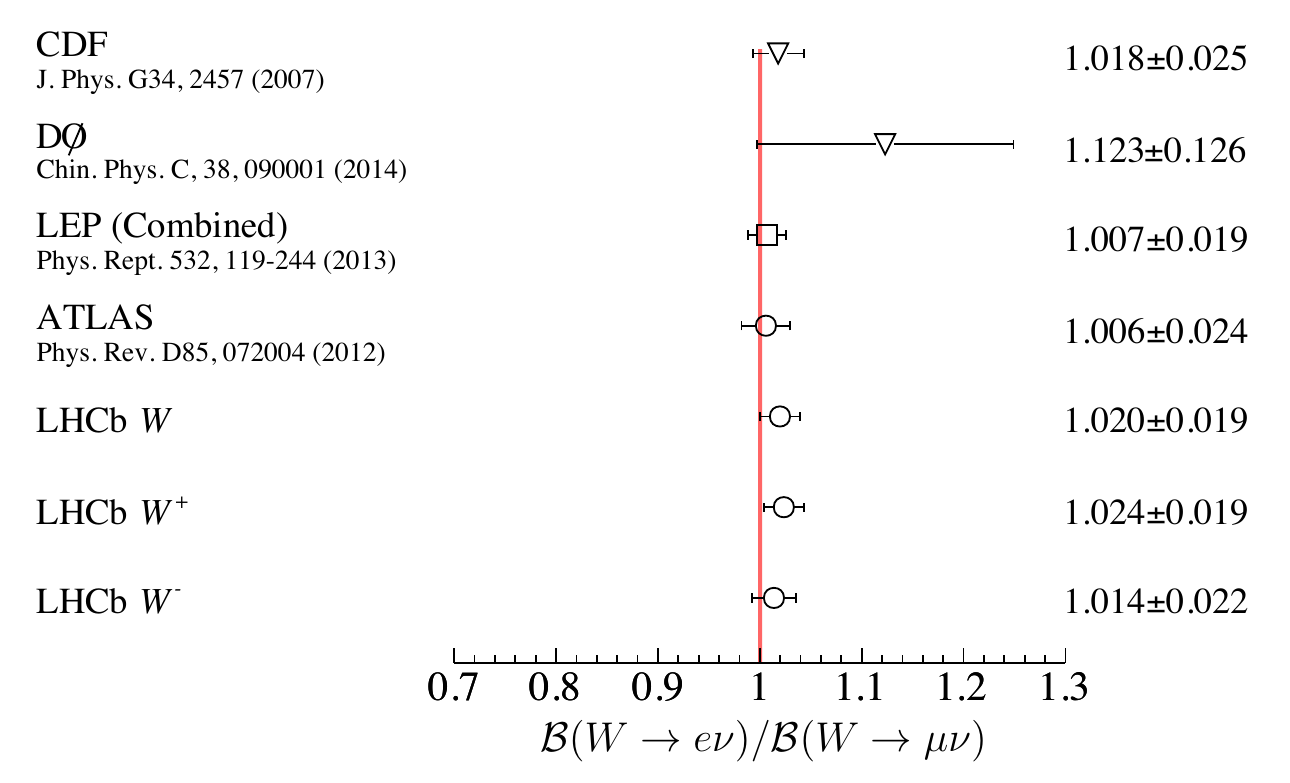}
\vspace*{0.5cm}
\vspace*{-0.5cm}
\end{center}
\caption{
	The ratio of branching fractions for the electron and muon final states determined for \W, \Wp, and \Wm is compared to hadron collider and LEP results. The theory expectation is represented by the red line.}
\label{fig:leptonUniversality}
\end{figure}
The consistency with lepton universality is quantified by computing a ratio of \W branching fractions using cross-sections determined in the range $2.00<\eta^{l}<3.50$. In this limited acceptance, the ratios of \W branching fractions are determined to be
\input{results/leptonUniversality.tex}
\noindent where the first uncertainties are statistical and the second are systematic. The result is compared to past measurements~\cite{WZATLAS,LUCDF,PDG,LEPCombo} in Fig.~\ref{fig:leptonUniversality} and its precision is seen to exceed previous individual determinations of the ratio and to be comparable to the combined LEP result.
\clearpage

%% file: results/WxSec.tex
\begin{align*}
\begin{split}
	\sigma_{\Wp\to\ep\nu_{\electron}}&=1124.4\pm 2.1\pm 21.5\pm 11.2\pm 13.0\pb,\\
	\sigma_{\Wm\to\electron^{-}\neub_{\electron}}&=\,\,\,809.0\pm 1.9\pm 18.1\pm\,\,\,7.0\pm \phantom{0}9.4\pb,\\
	\sigma_{\W\to\electron\nu}&=1933.3\pm 2.9\pm 38.2\pm 18.2\pm 22.4\pb,
\end{split}
\end{align*}

%% file: results/ratio.tex
\begin{align*}
\begin{split}
 R_{\W^{\pm}}=1.390\pm 0.004\pm 0.013\pm 0.002,
\end{split}
\end{align*}

%% file: results/leptonUniversality.tex
\begin{align*}
\begin{split}
	\BF(\Wp\to\ep\nu_{\electron})/\BF(\Wp\to\mup\nu_{\muon})&=1.024\pm 0.003\pm 0.019,\\
	\BF(\Wm\to\electron^{-}\neub_{\electron})/\BF(\Wm\to\mun\neub_{\muon})&=1.014\pm 0.004\pm 0.022,\\
	\BF(\W\to\electron\nu)/\BF(\W\to\muon\nu)\,\,\,\,\,\,\,\,\,\,&=1.020\pm 0.002\pm 0.019,
\end{split}
\end{align*}

%% file: conclusions.tex
\section{Conclusions}
\label{sec:conclusions}

 Measurements of the cross-sections for \W boson production in $pp$ collisions are presented at a centre-of-mass energy of $\sqrt{s}=8$\tev in the electron final state. The cross-section ratio and the charge asymmetry are also determined. The measurements are found to be in agreement with NNLO calculations in perturbative QCD.

These results represent the first measurements of $\W\to\electron\nu$ production in the forward region at the LHC and are complementary to the previously published measurements of $\W\to\mu\nu$ production. The measurements have been performed using statistically independent datasets with largely independent systematic uncertainties. The measurements reported here are found to be consistent with the $\W\to\mu\nu$ results.

Comparable precision to the $\W\to\mu\nu$ results is achieved in the measurements of the cross-sections and the cross-section ratio has been determined with sub-percent precision. Due to the unique kinematic acceptance of the \lhcb detector these results will be valuable in constraining the parton distribution functions of the proton at low and high values of the Bjorken-$x$ variable.

Finally, the measurements of \W production in the electron and muon final states are consistent with lepton universality and the ratio of branching fractions has precision that exceeds all past determinations at hadron colliders as well as measurements made at the LEP collider.

%% file: acknowledgements.tex
\section*{Acknowledgements}
\noindent We express our gratitude to our colleagues in the CERN
accelerator departments for the excellent performance of the LHC. We
thank the technical and administrative staff at the LHCb
institutes. We acknowledge support from CERN and from the national
agencies: CAPES, CNPq, FAPERJ and FINEP (Brazil); NSFC (China);
CNRS/IN2P3 (France); BMBF, DFG and MPG (Germany); INFN (Italy); 
FOM and NWO (The Netherlands); MNiSW and NCN (Poland); MEN/IFA (Romania); 
MinES and FANO (Russia); MinECo (Spain); SNSF and SER (Switzerland); 
NASU (Ukraine); STFC (United Kingdom); NSF (USA).
We acknowledge the computing resources that are provided by CERN, IN2P3 (France), KIT and DESY (Germany), INFN (Italy), SURF (The Netherlands), PIC (Spain), GridPP (United Kingdom), RRCKI and Yandex LLC (Russia), CSCS (Switzerland), IFIN-HH (Romania), CBPF (Brazil), PL-GRID (Poland) and OSC (USA). We are indebted to the communities behind the multiple open 
source software packages on which we depend.
Individual groups or members have received support from AvH Foundation (Germany),
EPLANET, Marie Sk\l{}odowska-Curie Actions and ERC (European Union), 
Conseil G\'{e}n\'{e}ral de Haute-Savoie, Labex ENIGMASS and OCEVU, 
R\'{e}gion Auvergne (France), RFBR and Yandex LLC (Russia), GVA, XuntaGal and GENCAT (Spain), Herchel Smith Fund, The Royal Society, Royal Commission for the Exhibition of 1851 and the Leverhulme Trust (United Kingdom).

%% file: appendix.tex
\clearpage
{\noindent\bf\Large Appendices}
\appendix

\section{Tabulated results}
\label{sec:tabulatedResults}

Born level cross-sections in bins of electron pseudorapidity for $\Wp$ ($\Wm$) along with corresponding FSR corrections are given in Table~\ref{tab:WpxSecs} (\ref{tab:WmxSecs}). The ratio is given in Table~\ref{tab:ratio} and the charge asymmetry in Table~\ref{tab:Asy}.
\input{results/Table2.tex}
\input{results/Table3.tex}
\input{results/Table4.tex}
\input{results/Table5.tex}

\clearpage

\section{Correlation coefficients}
\label{sec:correlationCoefficients}

The correlation coefficients of the systematic uncertainties between bins of \etae for the $\Wp$ ($\Wm$) cross-sections are given in Table~\ref{tab:Wp} (\ref{tab:Wm}) while those between bins for $\Wp$ and $\Wm$ are given in Table~\ref{tab:WpvWm}. The LHC beam energy and luminosity uncertainties, which are fully correlated between cross-section measurements, are excluded.
\input{results/Table6.tex}
\vspace{-15pt}
\input{results/Table7.tex}
\vspace{-15pt}
\input{results/Table8.tex}
\clearpage
\section{Fits to lepton \pt}
\label{sec:leptonptFits}

\begin{figure}[h]
\begin{center}
\begin{minipage}{.5\textwidth}
	\centering
		\includegraphics[width=1.05\linewidth]{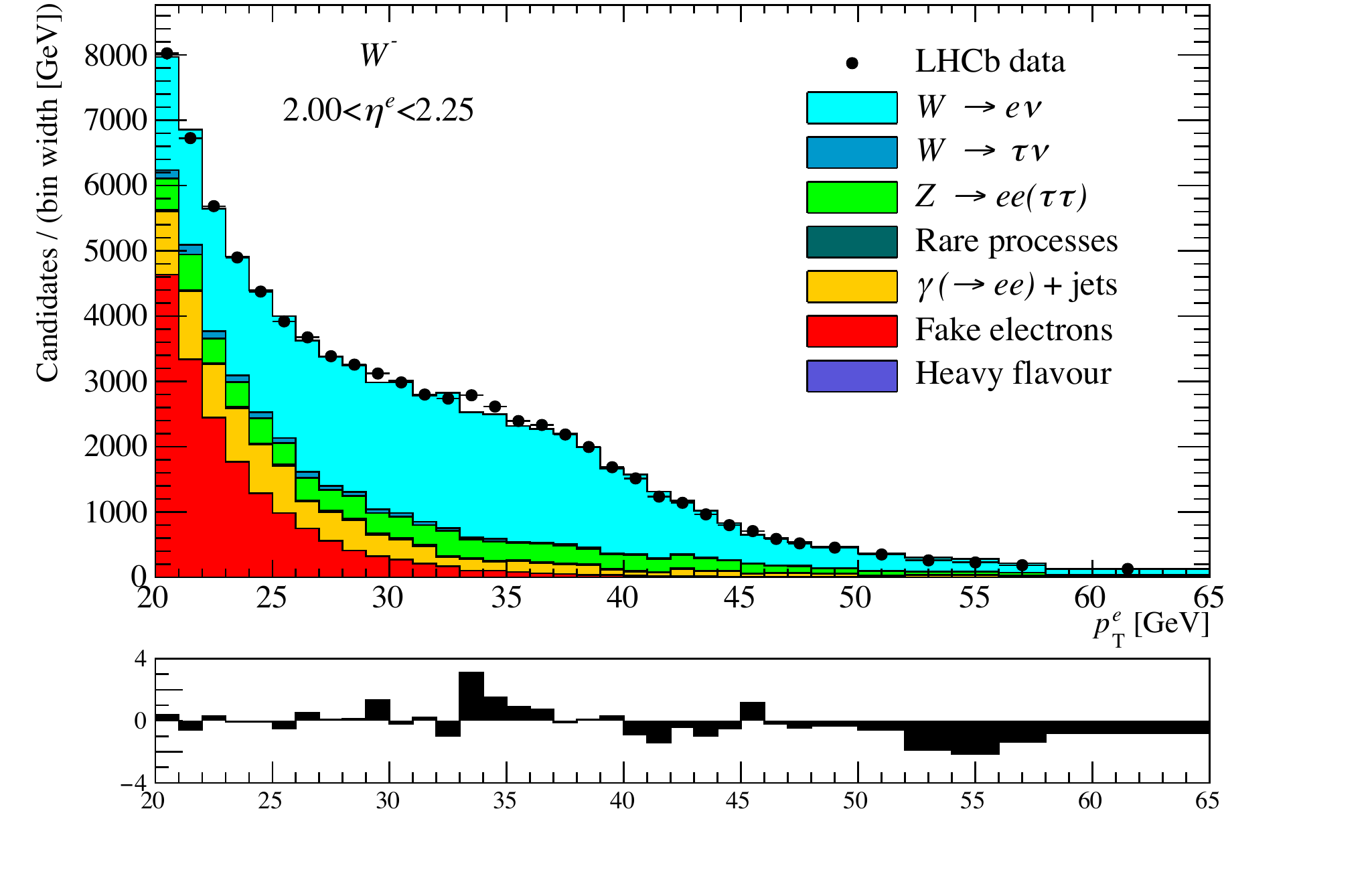}
\end{minipage}
\hspace{-0.25cm}
\begin{minipage}{0.5\textwidth}
	\centering
		\includegraphics[width=1.05\linewidth]{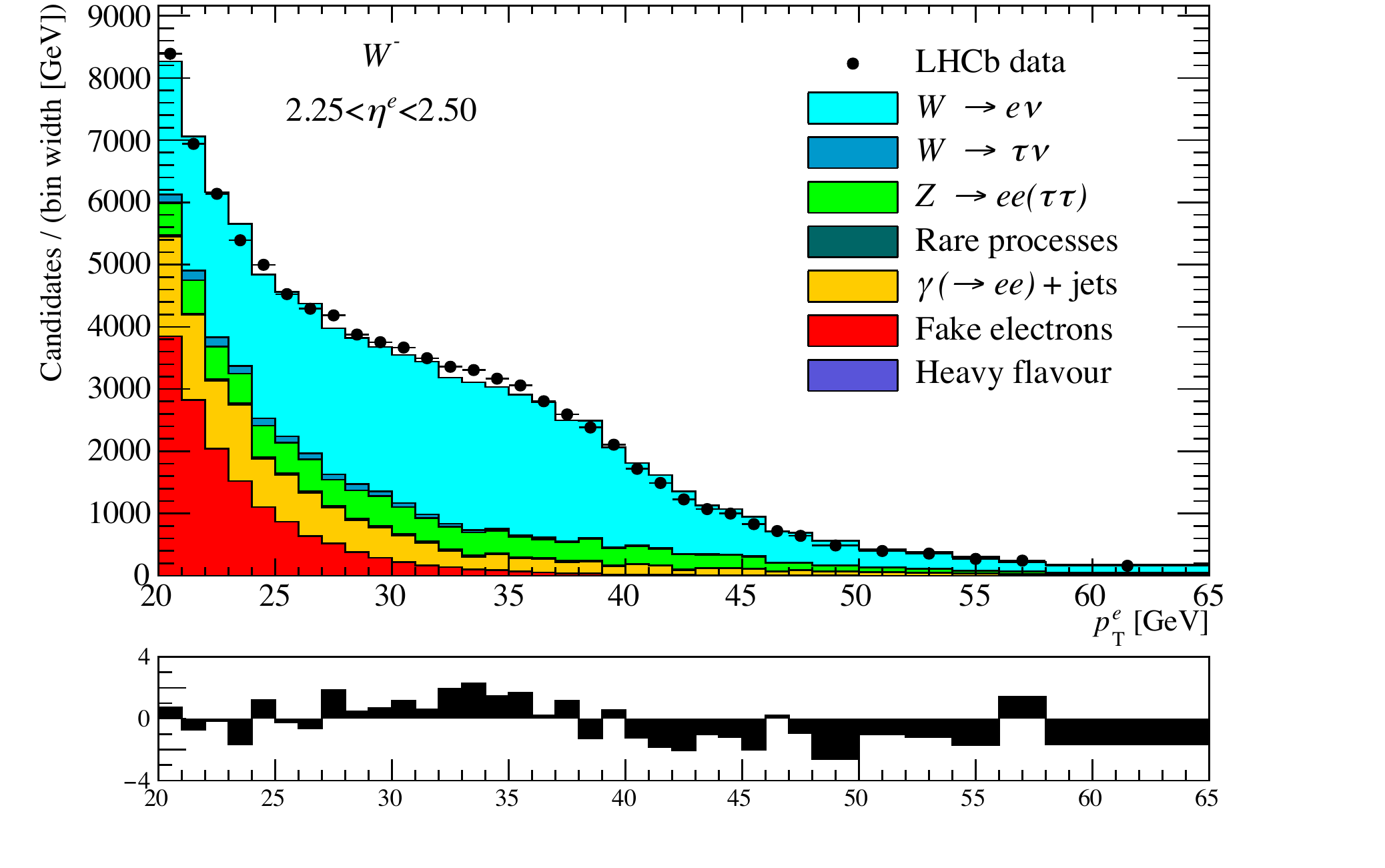}	
\end{minipage}
\begin{minipage}{.5\textwidth}
	\centering
		\includegraphics[width=1.05\linewidth]{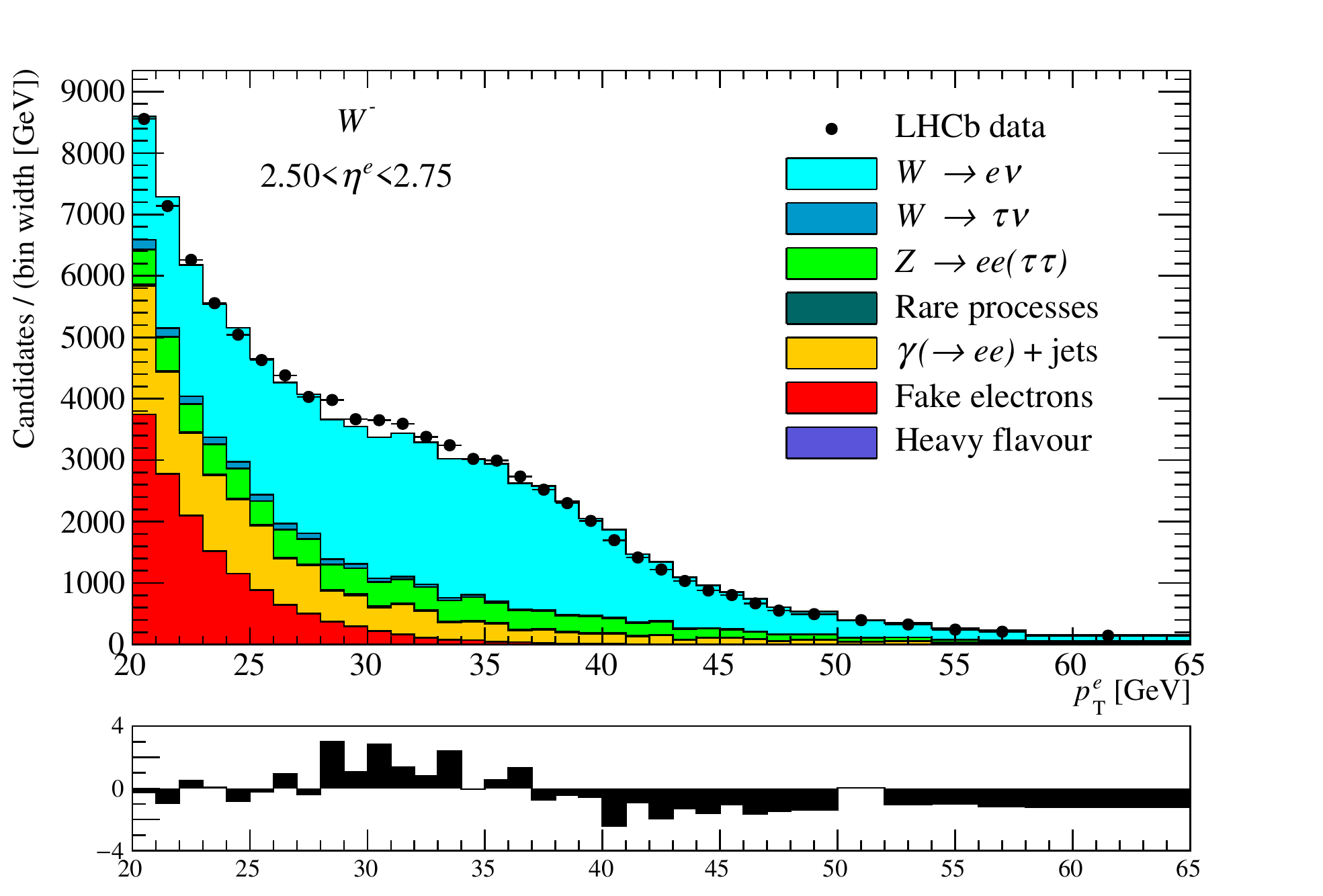}
\end{minipage}
\hspace{-0.25cm}
\begin{minipage}{0.5\textwidth}
	\centering
		\includegraphics[width=1.05\linewidth]{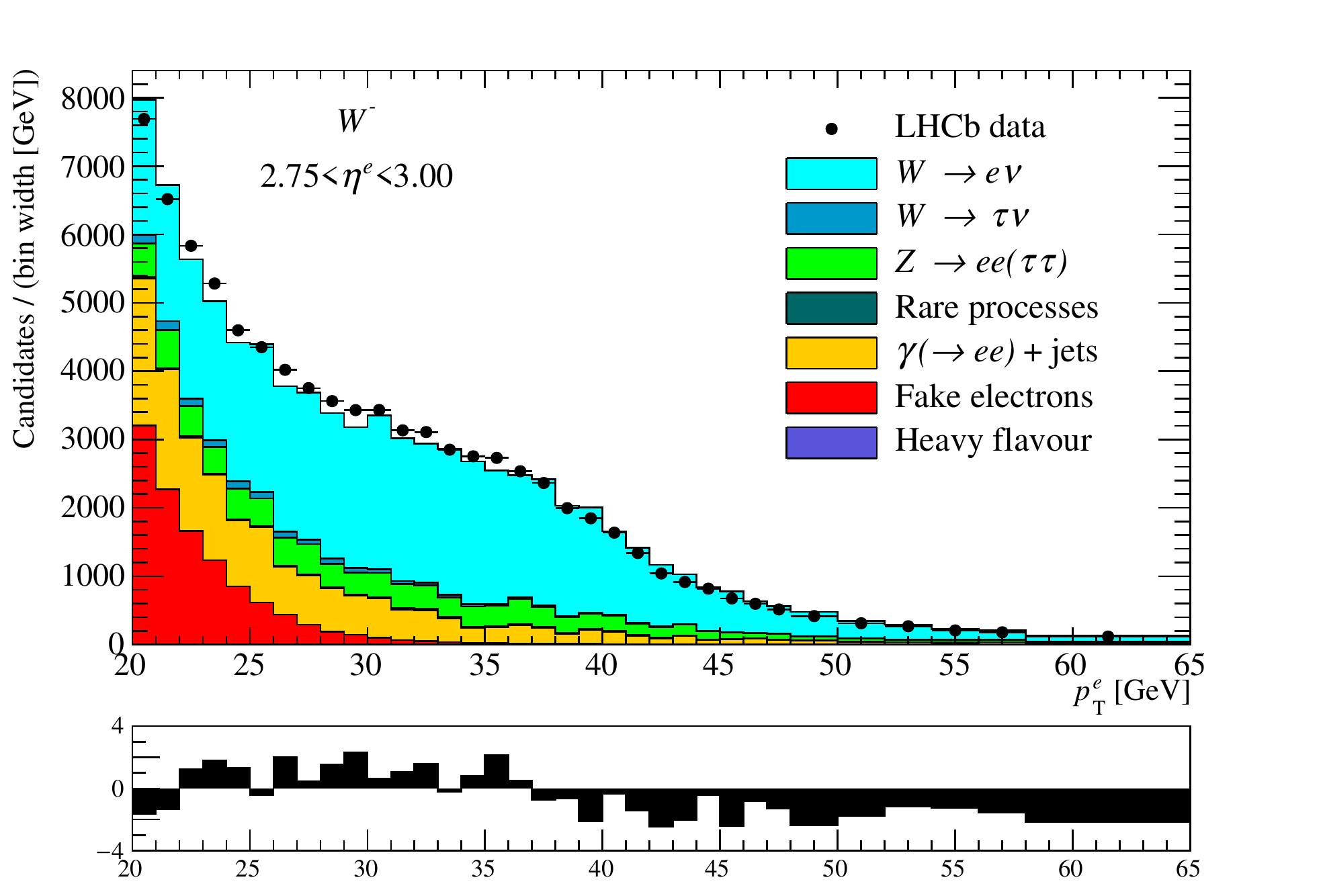}	
\end{minipage}
\begin{minipage}{.5\textwidth}
	\centering
		\includegraphics[width=1.05\linewidth]{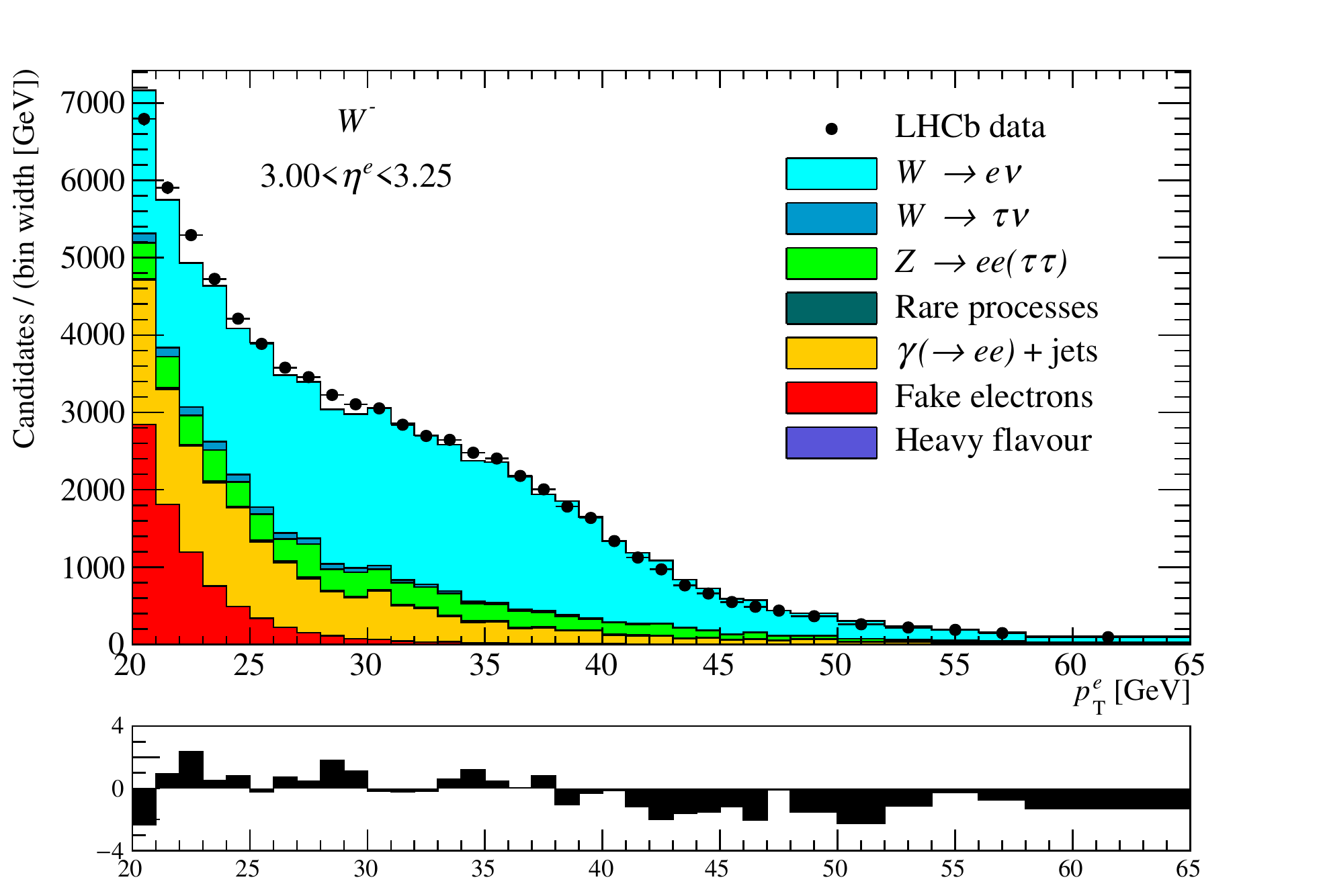}
\end{minipage}
\hspace{-0.25cm}
\begin{minipage}{0.5\textwidth}
	\centering
		\includegraphics[width=1.05\linewidth]{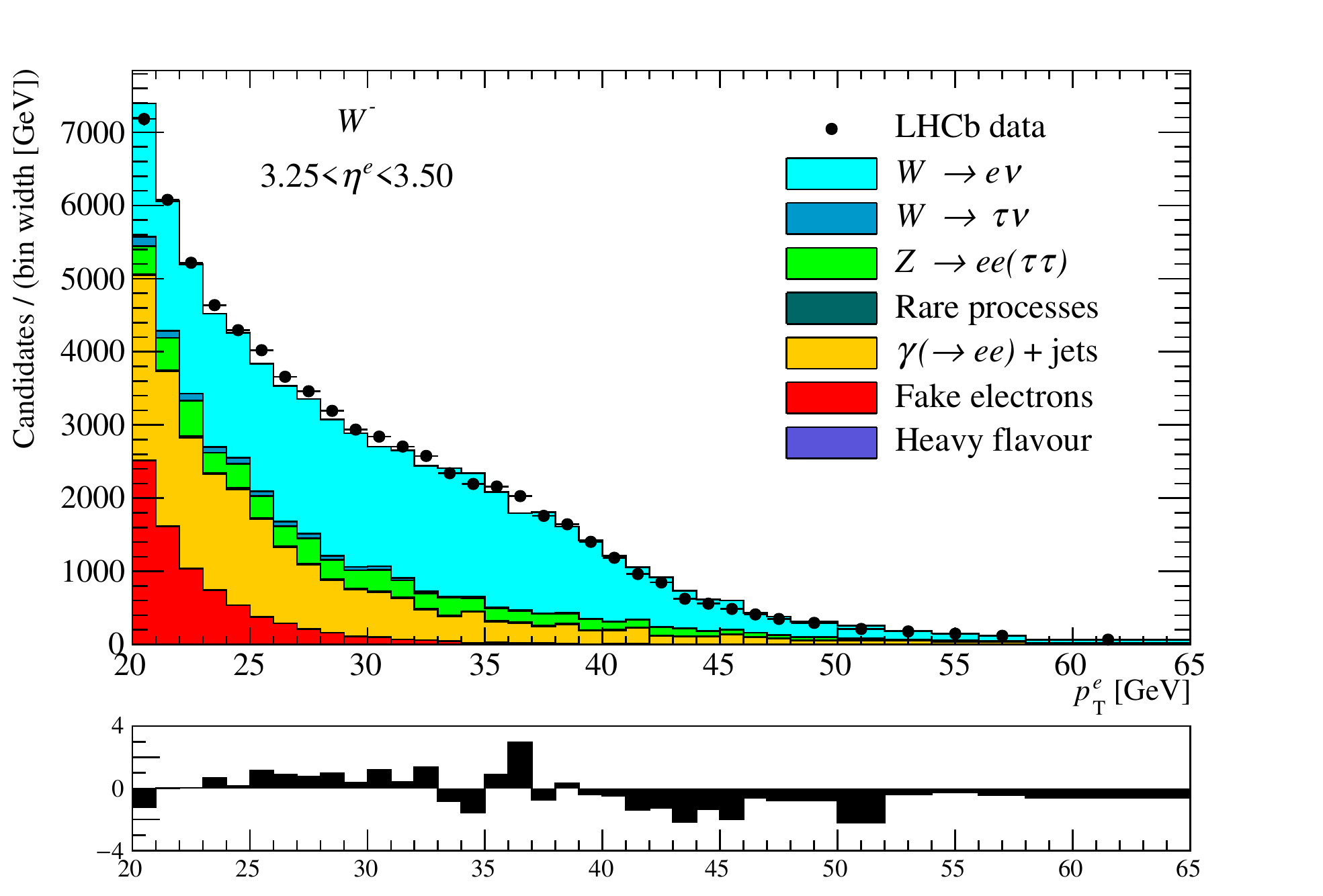}	
\end{minipage}
\begin{minipage}{.5\textwidth}
	\centering
		\includegraphics[width=1.05\linewidth]{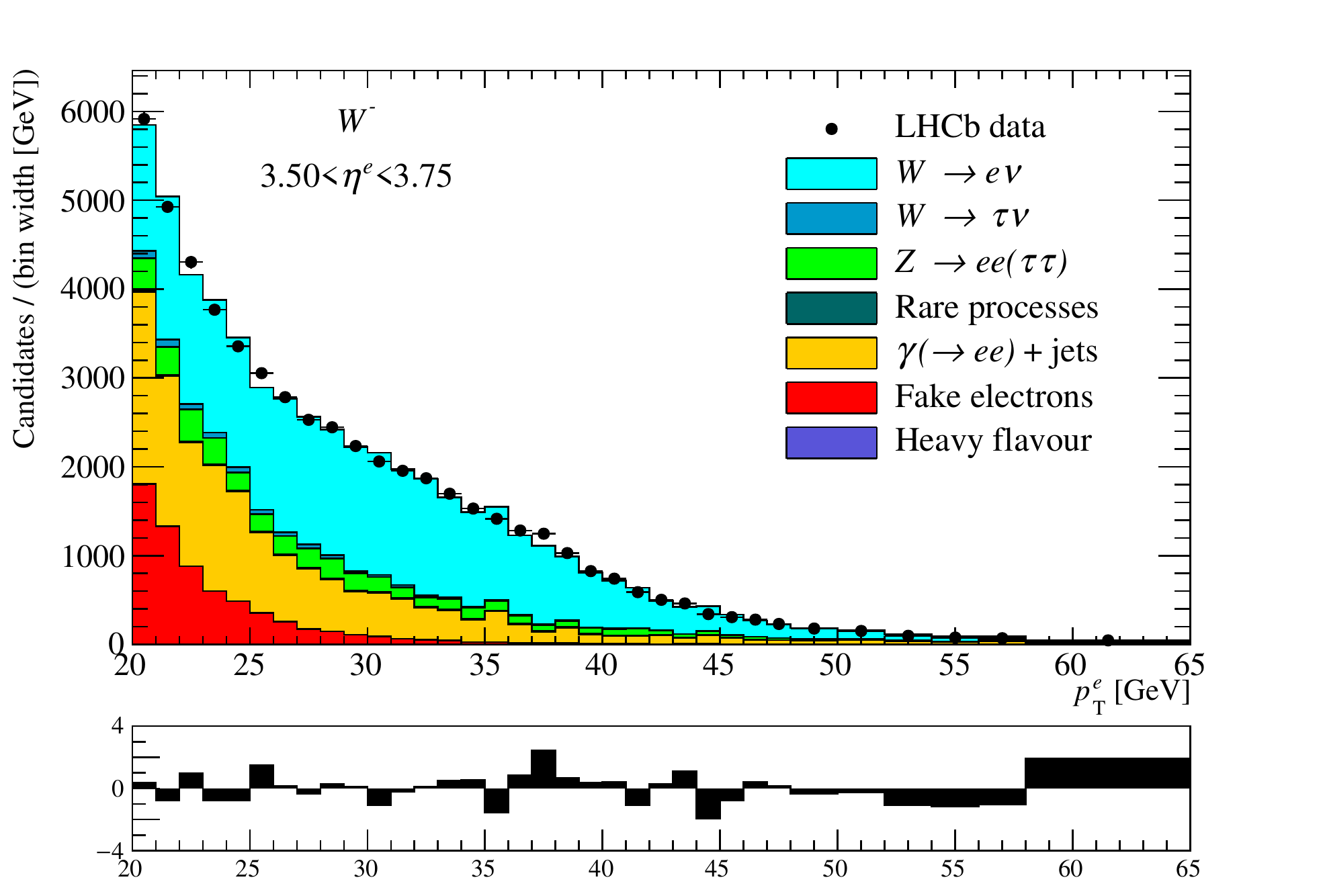}
\end{minipage}
\hspace{-0.25cm}
\begin{minipage}{0.5\textwidth}
	\centering
		\includegraphics[width=1.05\linewidth]{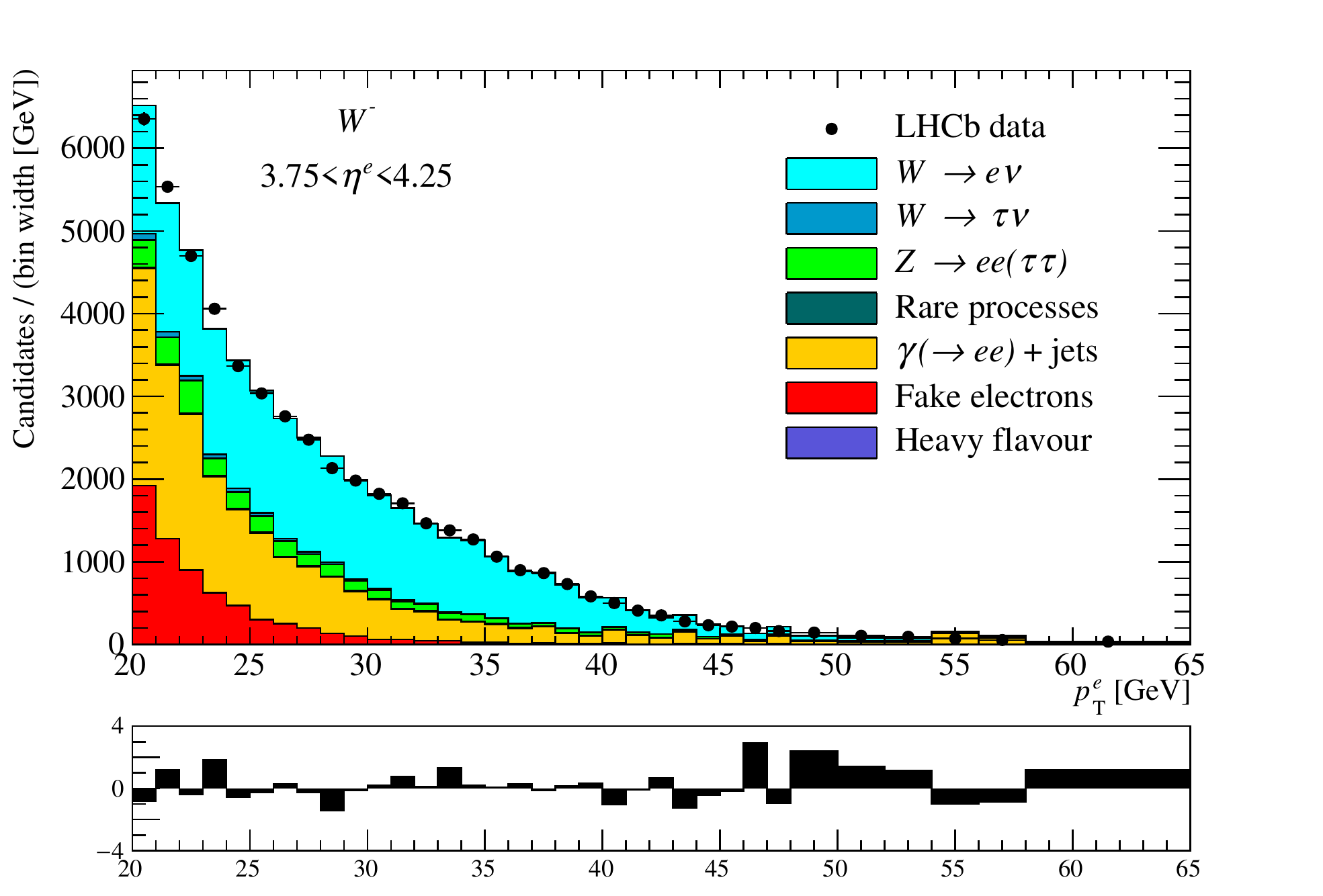}	
\end{minipage}
\end{center}
\caption{Fits to \pte for \en in bins of $\eta^{\electron}$. Pulls are shown underneath.}
\label{fig:binnedpTFitsel_linear}
\end{figure}

\begin{figure}[tb]
\begin{center}
\begin{minipage}{.5\textwidth}
	\centering
		\includegraphics[width=1.05\linewidth]{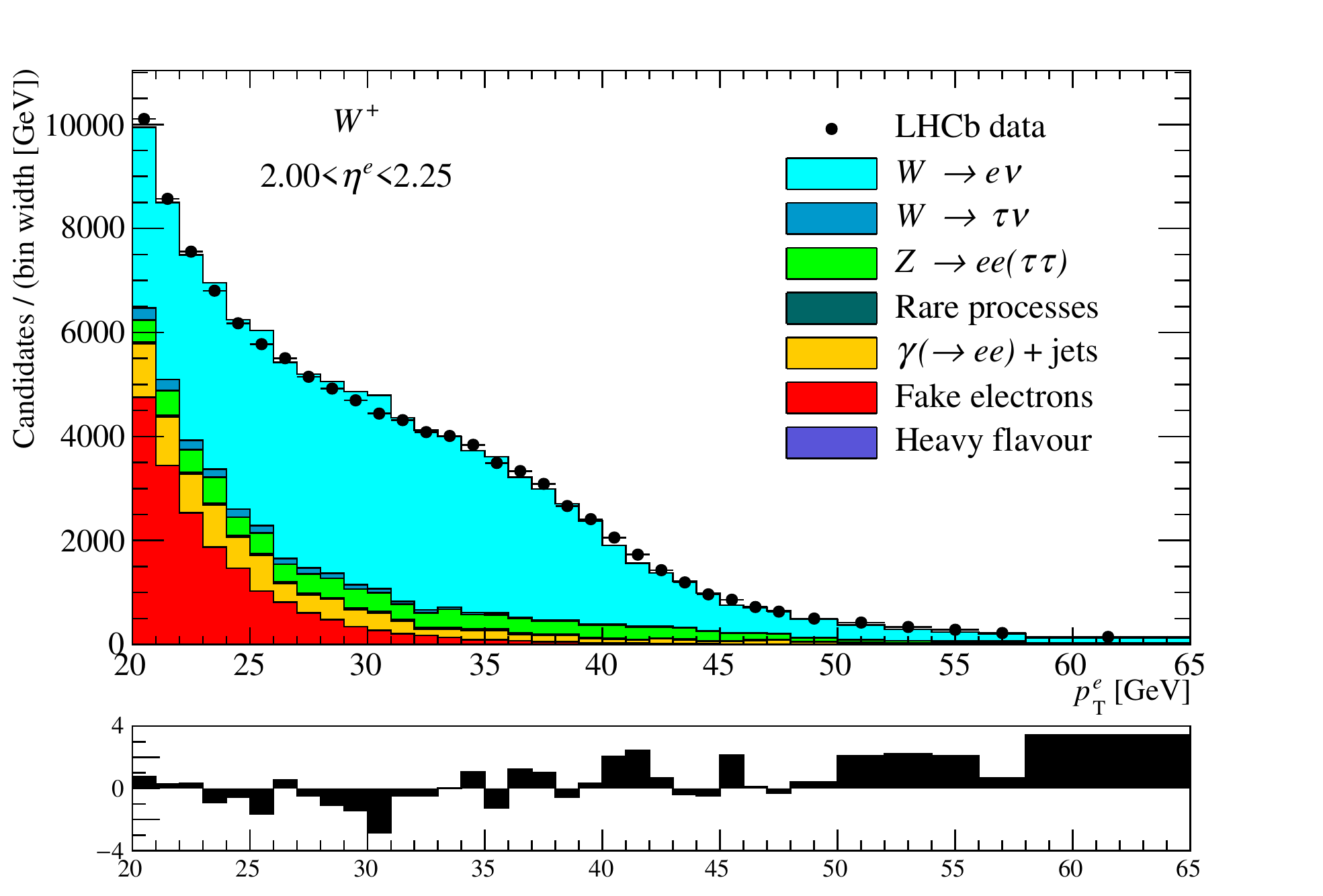}
\end{minipage}
\hspace{-0.25cm}
\begin{minipage}{0.5\textwidth}
	\centering
		\includegraphics[width=1.05\linewidth]{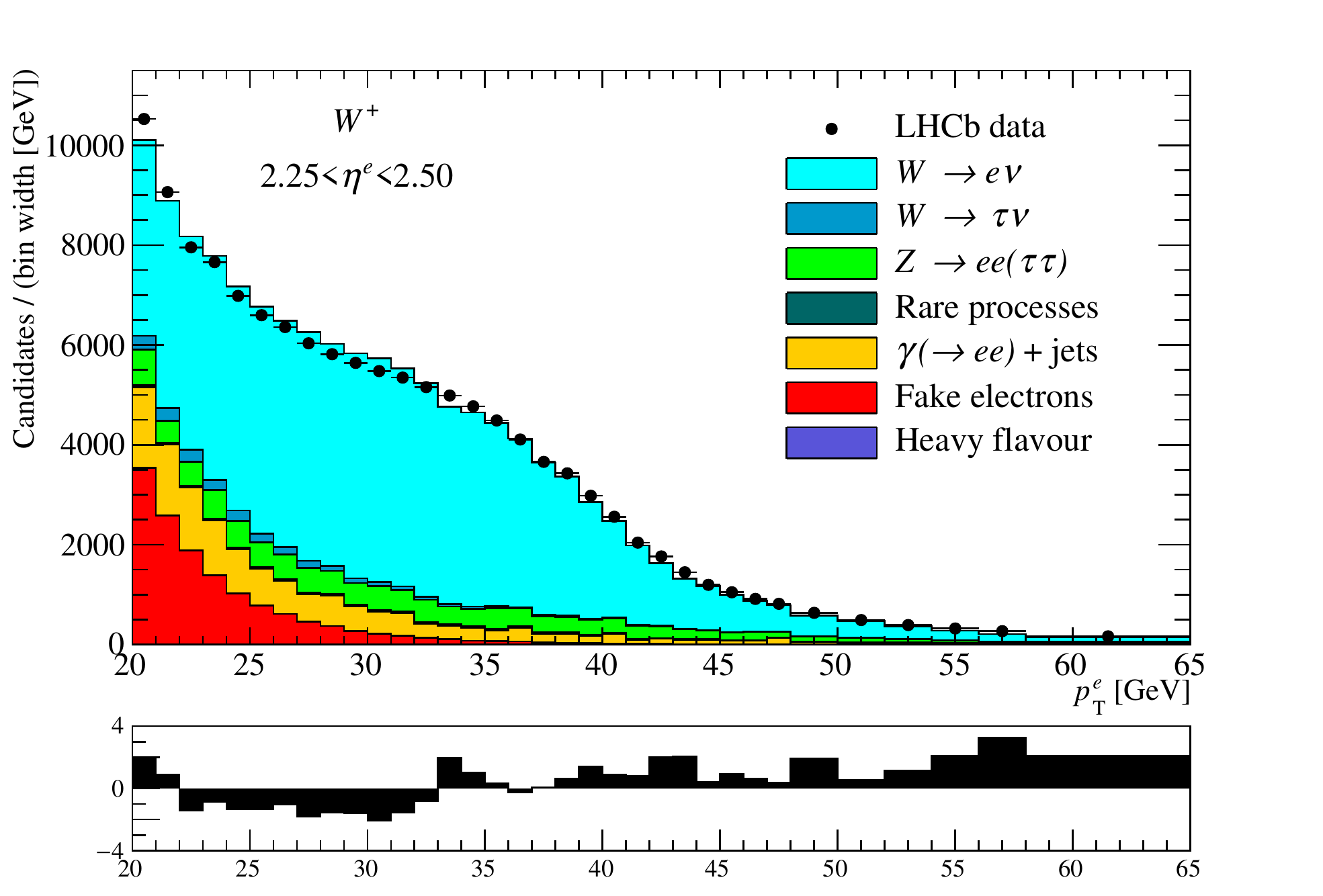}	
\end{minipage}
\begin{minipage}{.5\textwidth}
	\centering
		\includegraphics[width=1.05\linewidth]{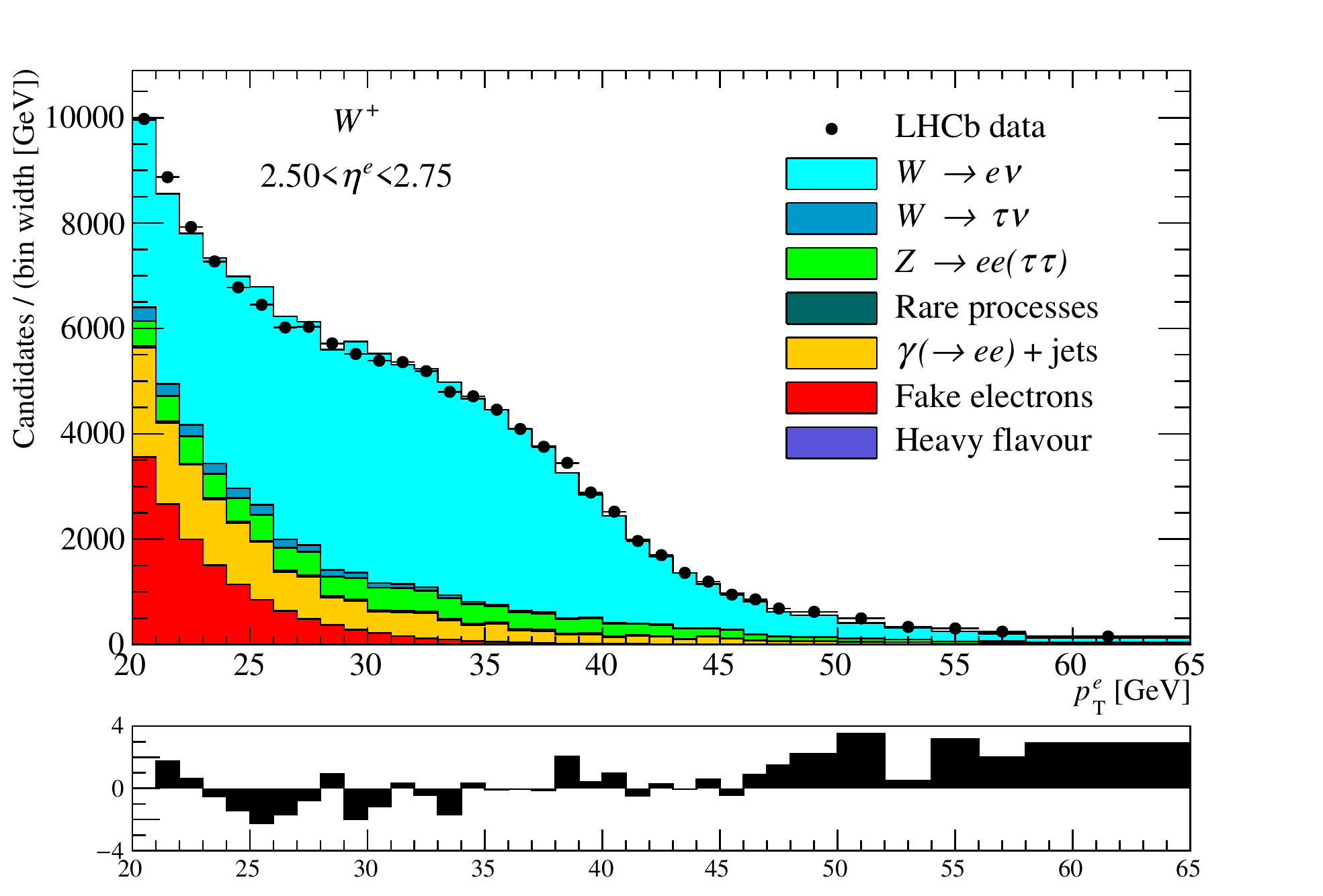}
\end{minipage}
\hspace{-0.25cm}
\begin{minipage}{0.5\textwidth}
	\centering
		\includegraphics[width=1.05\linewidth]{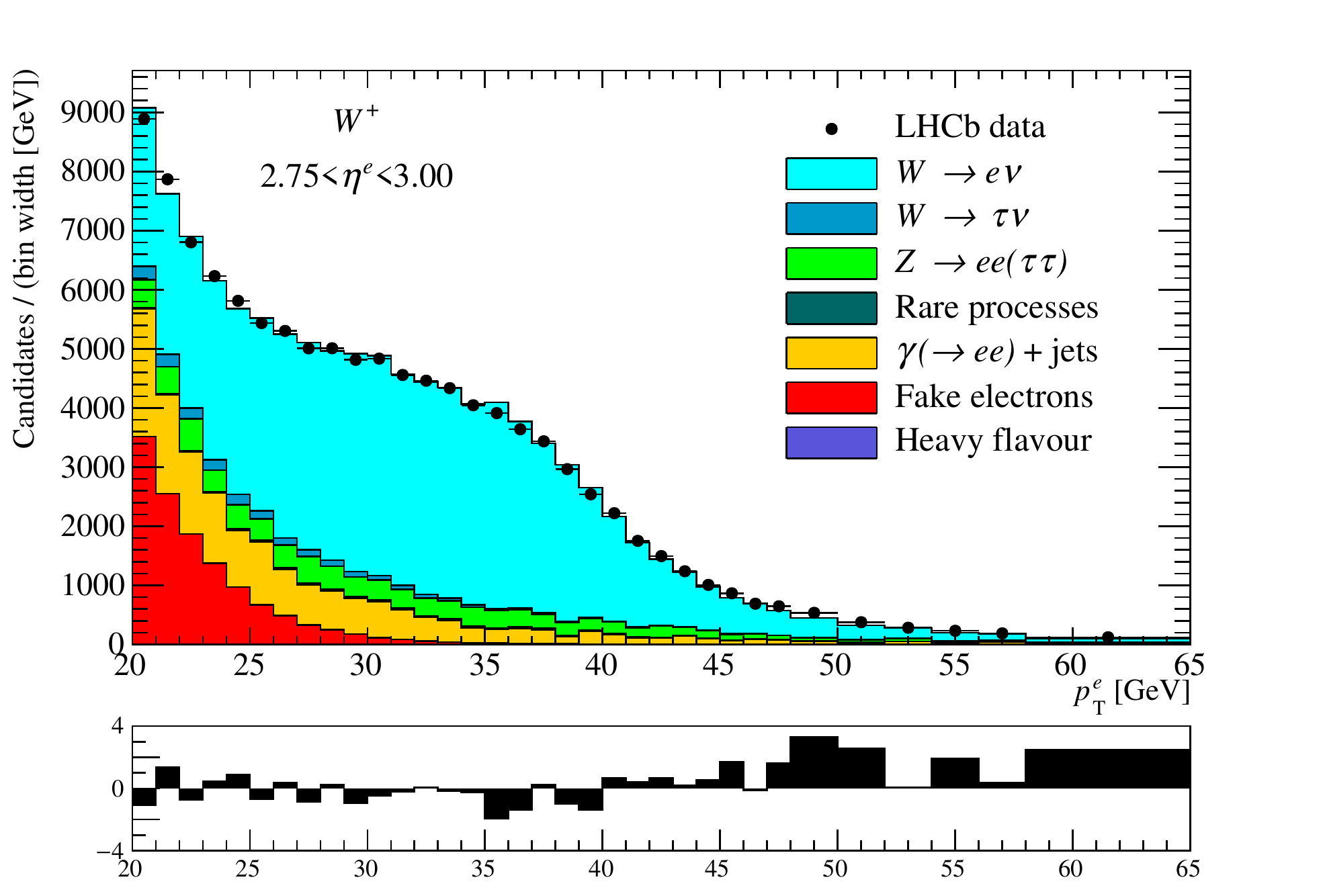}	
\end{minipage}
\begin{minipage}{.5\textwidth}
	\centering
		\includegraphics[width=1.05\linewidth]{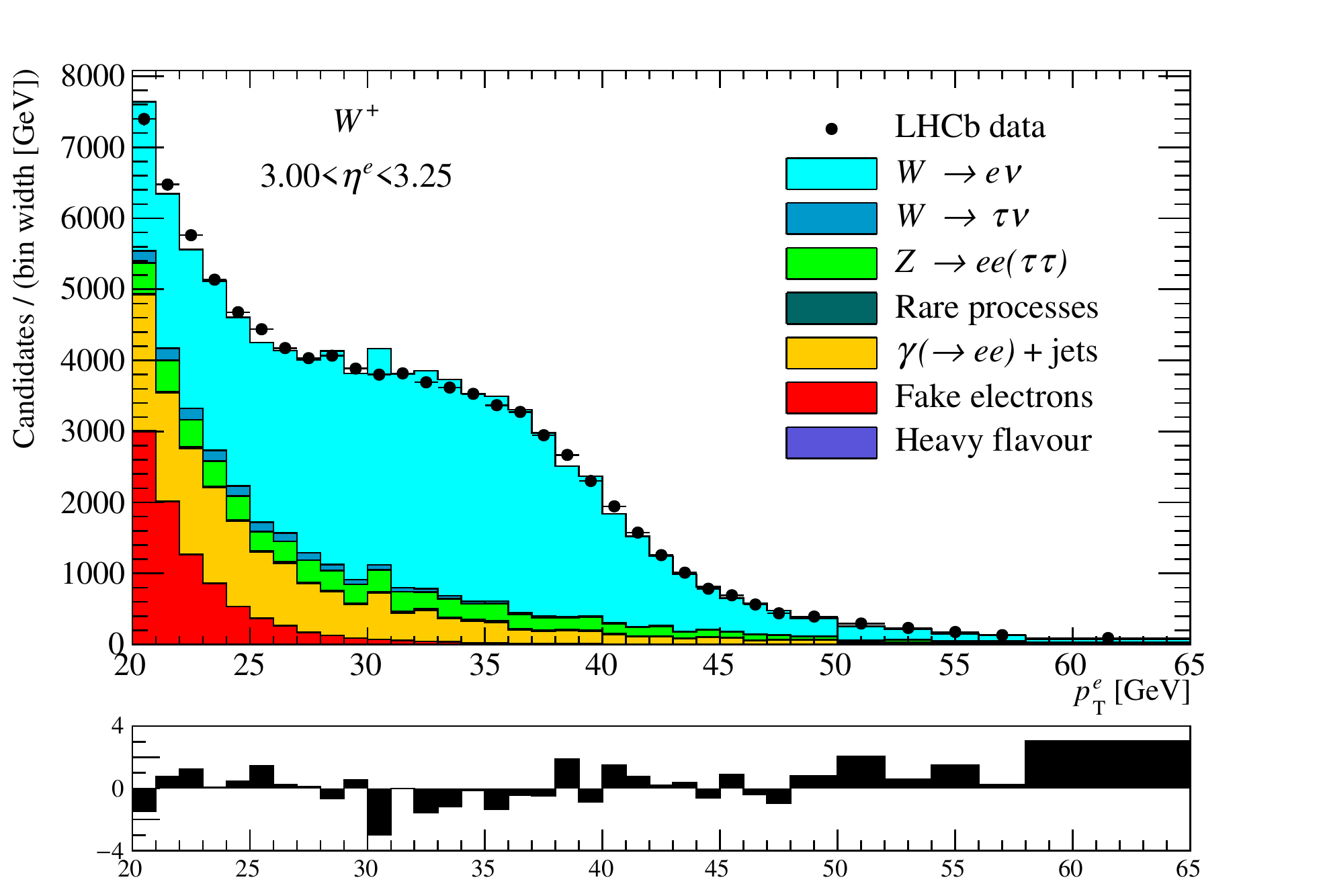}
\end{minipage}
\hspace{-0.25cm}
\begin{minipage}{0.5\textwidth}
	\centering
		\includegraphics[width=1.05\linewidth]{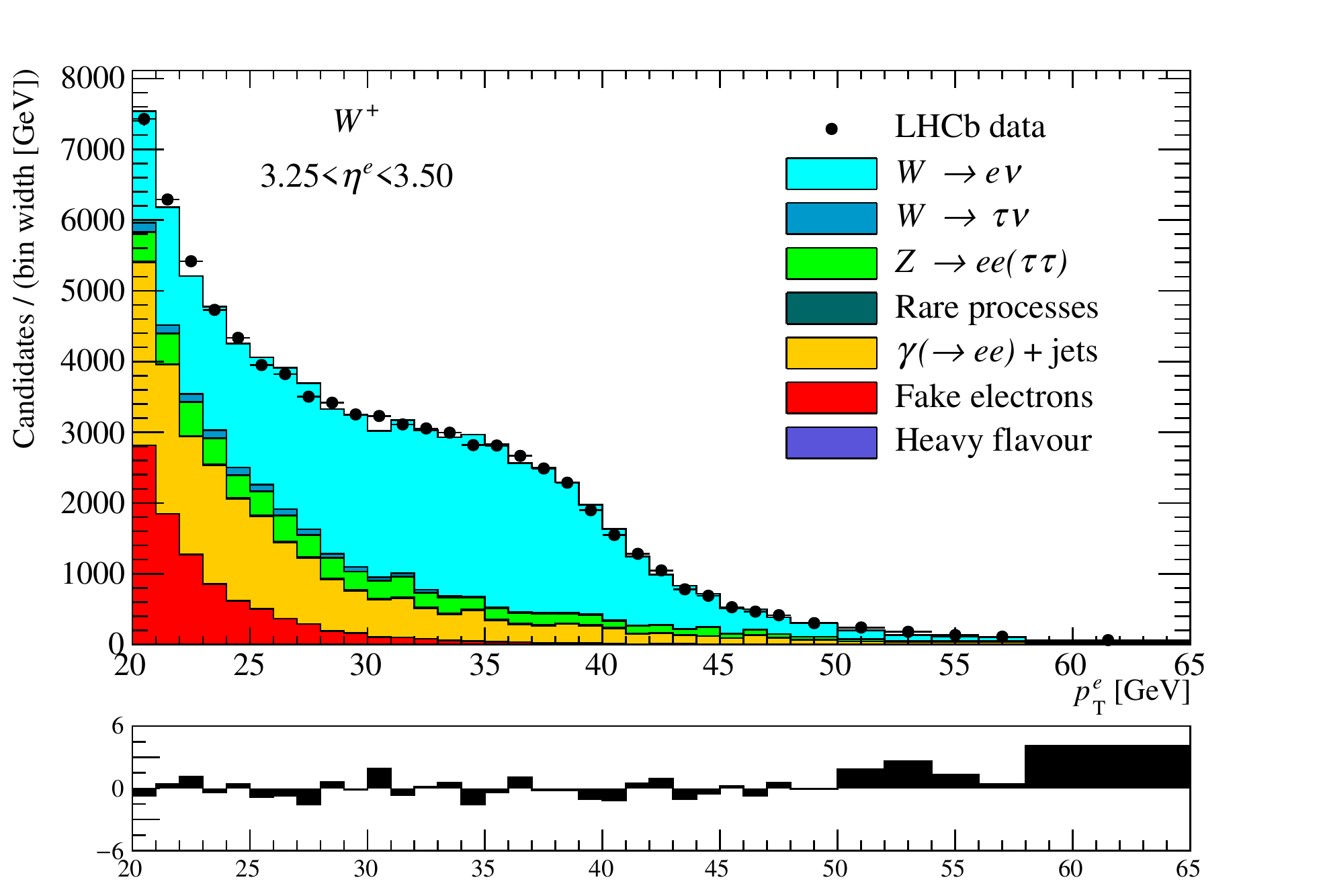}	
\end{minipage}
\begin{minipage}{.5\textwidth}
	\centering
		\includegraphics[width=1.05\linewidth]{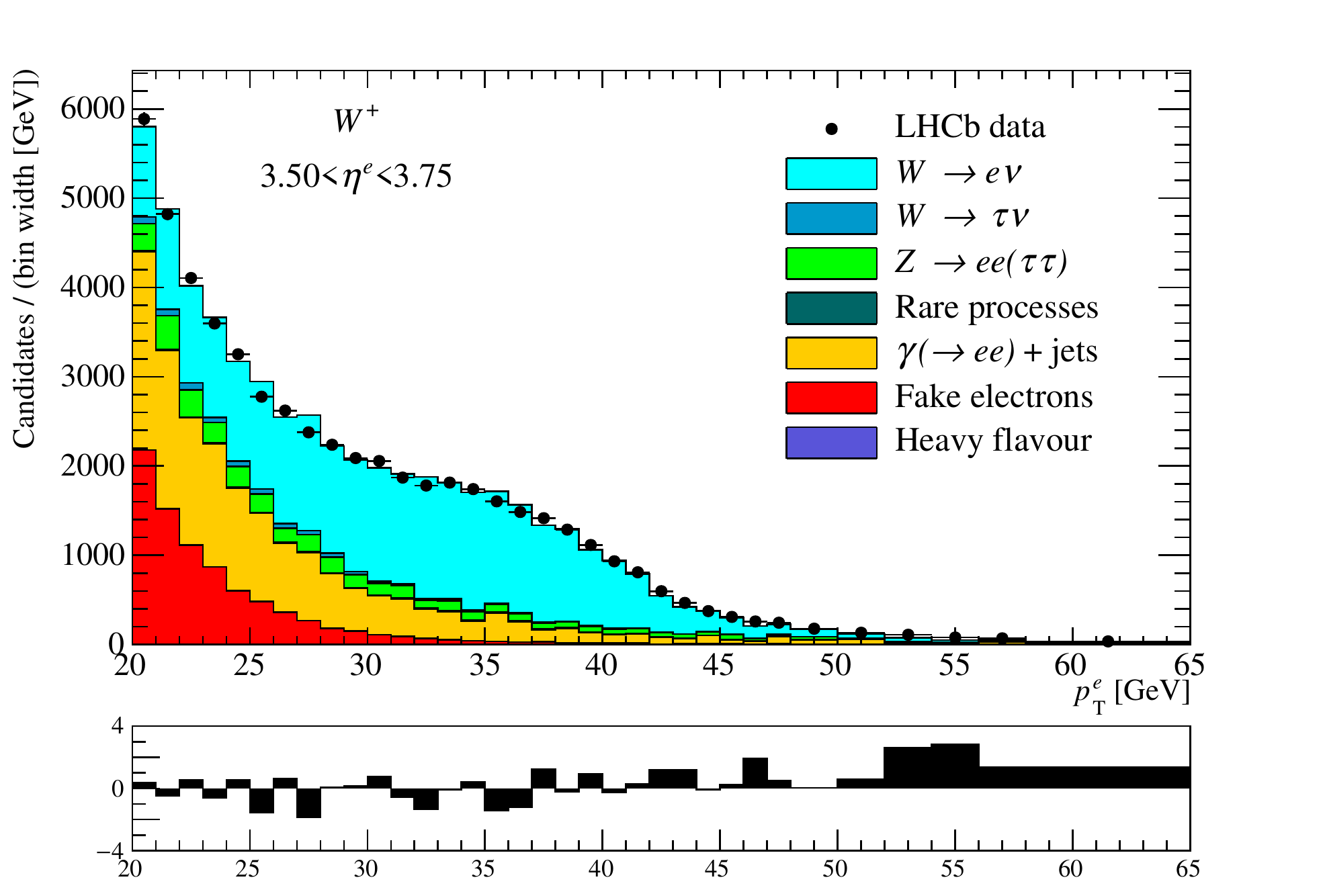}
\end{minipage}
\hspace{-0.25cm}
\begin{minipage}{0.5\textwidth}
	\centering
		\includegraphics[width=1.05\linewidth]{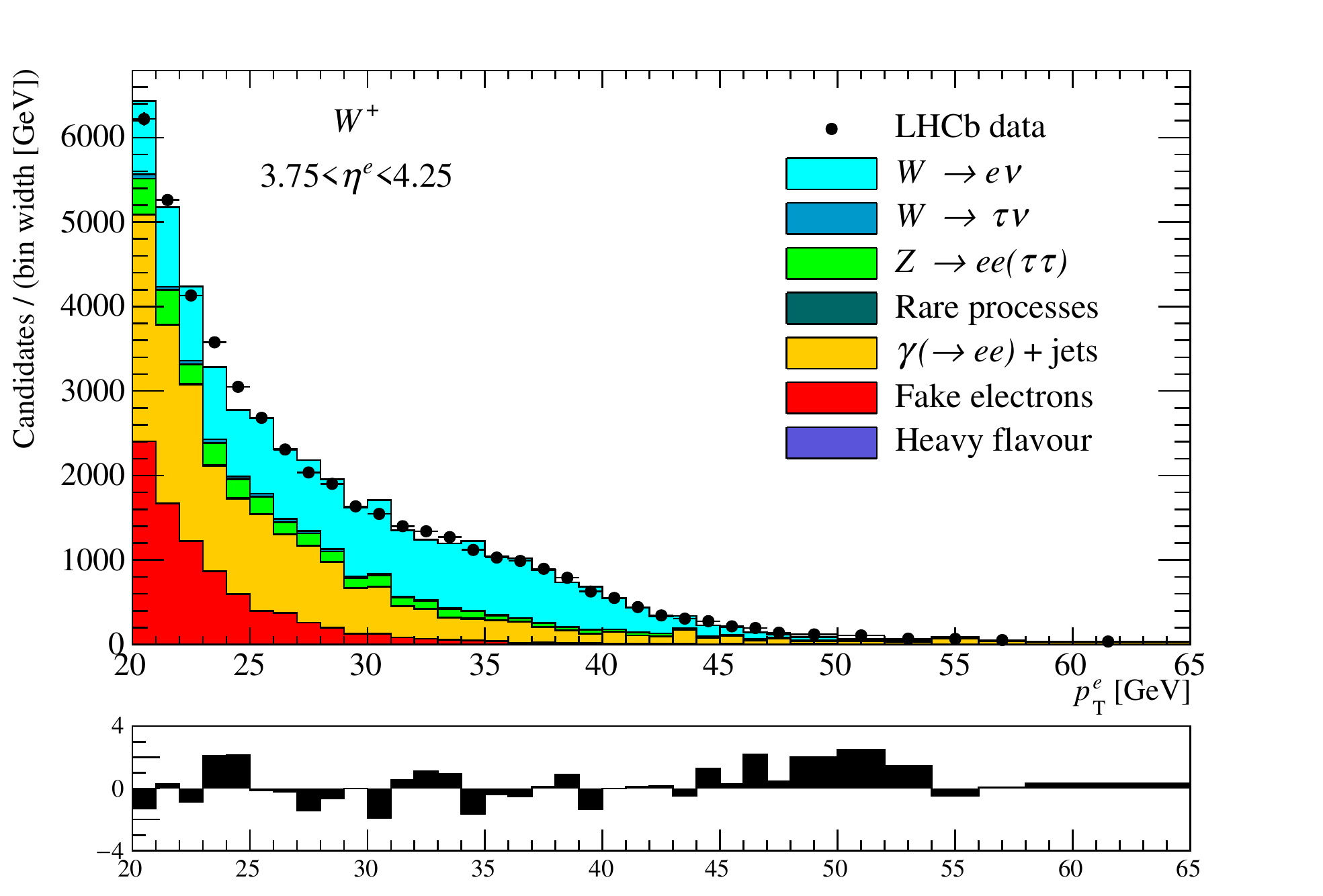}	
\end{minipage}
\end{center}
\caption{Fits to \pte for \ep in bins of $\eta^{\electron}$. Pulls are shown underneath.}
\label{fig:binnedpTFitsep_linear}
\end{figure}

The fits to \pte binned in $\eta^{\electron}$ are shown in Figs.~\ref{fig:binnedpTFitsel_linear} and~\ref{fig:binnedpTFitsep_linear}. The pulls shown underneath each fit are statistical only. The fractional signal contribution in the \Wp(\Wm) sample varies from \myTilde70\%(\myTilde60\%) near $\etae=2$ to \myTilde40\%(\myTilde50\%) at the largest pseudorapidity. The values of \chisqndf for the fits range between 0.9 and 2.3, based on statistical uncertainties only. The systematic uncertainties in the event yields presented in Section~\ref{sec:systematicUncertainties} are found to cover the uncertainty that arises from imperfect fit quality.

\clearpage

%% file: results/Table2.tex
\begin{table}[h]
\begin{center}
\begin{tabular}{ccc}
$\eta^{e}$ & $\sigma_{\W^{+}\to\ep\nu_{\electron}} [\pb]$ & $f^{\textup{FSR}}$ \\
\hline
2.00 - 2.25&$229.9\pm1.0\pm6.5\pm2.3\pm2.7$&$0.9671\pm0.0013$\\
2.25 - 2.50&$210.1\pm0.8\pm4.7\pm2.1\pm2.4$&$0.9714\pm0.0013$\\
2.50 - 2.75&$191.7\pm0.8\pm4.9\pm1.9\pm2.2$&$0.9718\pm0.0013$\\
2.75 - 3.00&$156.3\pm0.7\pm3.4\pm1.6\pm1.8$&$0.9741\pm0.0015$\\
3.00 - 3.25&$132.0\pm0.7\pm3.1\pm1.3\pm1.5$&$0.9739\pm0.0016$\\
3.25 - 3.50&\,\,\,$87.6\pm0.6\pm2.2\pm0.9\pm1.0$&$0.9697\pm0.0019$\\
3.50 - 3.75&\,\,\,$59.1\pm0.5\pm2.1\pm0.6\pm0.7$&$0.9727\pm0.0023$\\
3.75 - 4.25&\,\,\,$57.8\pm0.7\pm2.7\pm0.6\pm0.7$&$0.9672\pm0.0024$\\
\end{tabular}
\end{center}
\caption{The Born level cross-section for $W^{+}$ boson production in bins of electron pseudorapidity. The first uncertainties are statistical, the second are systematic, the third are due to the knowledge of the LHC beam energy and the fourth are due to the luminosity measurement.  The rightmost column gives values of the additional factor, $f^{\textup{FSR}}$, by which the results should be multiplied in order to give the cross-sections after FSR.}
\label{tab:WpxSecs}
\end{table}

%% file: results/Table3.tex
\begin{table}[h]
\begin{center}
\begin{tabular}{ccc}
$\eta^{e}$ & $\sigma_{\W^{-}\to\electron^{-}\neub_{\electron}} [\pb]$ & $f^{\textup{FSR}}$ \\
\hline
2.00 - 2.25&$132.8\pm0.8\pm4.1\pm1.1\pm1.5$&$0.9729\pm0.0021$\\
2.25 - 2.50&$120.8\pm0.7\pm3.1\pm1.0\pm1.4$&$0.9726\pm0.0020$\\
2.50 - 2.75&$113.0\pm0.7\pm2.9\pm1.0\pm1.3$&$0.9762\pm0.0020$\\
2.75 - 3.00&$103.3\pm0.6\pm2.7\pm0.9\pm1.2$&$0.9786\pm0.0019$\\
3.00 - 3.25&\,\,\,$99.3\pm0.6\pm2.7\pm0.9\pm1.2$&$0.9746\pm0.0019$\\
3.25 - 3.50&\,\,\,$78.8\pm0.6\pm2.2\pm0.7\pm0.9$&$0.9756\pm0.0019$\\
3.50 - 3.75&\,\,\,$67.0\pm0.6\pm2.8\pm0.6\pm0.8$&$0.9713\pm0.0020$\\
3.75 - 4.25&\,\,\,$94.0\pm0.9\pm4.2\pm0.8\pm1.1$&$0.9653\pm0.0016$\\
\end{tabular}
\end{center}
\caption{The Born level cross-section for $W^{-}$ boson production in bins of electron pseudorapidity. The first uncertainties are statistical, the second are systematic, the third are due to the knowledge of the LHC beam energy and the fourth are due to the luminosity measurement.  The rightmost column gives values of the additional factor, $f^{\textup{FSR}}$, by which the results should be multiplied in order to give the cross-sections after FSR.}
\label{tab:WmxSecs}
\end{table}

%% file: results/Table4.tex
\begin{table}[h]
\begin{center}
\begin{tabular}{cc}
$\eta^{e}$ & $R_{W^{\pm}}$ \\
\hline
2.00 - 2.25&$1.731\pm0.013\pm0.026\pm0.003$\\
2.25 - 2.50&$1.739\pm0.012\pm0.025\pm0.003$\\
2.50 - 2.75&$1.697\pm0.012\pm0.022\pm0.003$\\
2.75 - 3.00&$1.512\pm0.011\pm0.023\pm0.002$\\
3.00 - 3.25&$1.330\pm0.011\pm0.019\pm0.002$\\
3.25 - 3.50&$1.111\pm0.010\pm0.025\pm0.002$\\
3.50 - 3.75&$0.882\pm0.011\pm0.023\pm0.001$\\
3.75 - 4.25&$0.615\pm0.010\pm0.022\pm0.001$\\
\end{tabular}
\end{center}
\caption{The $W^{+}$ to $W^{-}$ cross-section ratio in bins of electron pseudorapidity. The first uncertainties are statistical, the second are systematic and the third are due to the knowledge of the LHC beam energy.}
\label{tab:ratio}
\end{table}

%% file: results/Table5.tex
\begin{table}[h]
\begin{center}
\begin{tabular}{cc}
$\eta^{e}$ & $A_{e} (\%)$ \\
\hline
2.00 - 2.25&$\,\,\,\,\,26.78\pm0.36\pm0.70\pm0.07$\\
2.25 - 2.50&$\,\,\,\,\,26.98\pm0.32\pm0.66\pm0.07$\\
2.50 - 2.75&$\,\,\,\,\,25.84\pm0.33\pm0.60\pm0.07$\\
2.75 - 3.00&$\,\,\,\,\,20.39\pm0.36\pm0.74\pm0.07$\\
3.00 - 3.25&$\,\,\,\,\,14.15\pm0.39\pm0.70\pm0.07$\\
3.25 - 3.50&$\,\,\,\,\,\,\,\,5.25\pm0.47\pm1.11\pm0.07$\\
3.50 - 3.75&$\,\,\,-6.25\pm0.60\pm1.28\pm0.07$\\
3.75 - 4.25&$-23.85\pm0.75\pm1.72\pm0.07$\\
\end{tabular}
\end{center}
\caption{The \W boson production charge asymmetry in bins of electron pseudorapidity. The first uncertainties are statistical, the second are systematic and the third are due to the knowledge of the LHC beam energy.}
\label{tab:Asy}
\end{table}

%% file: results/Table6.tex
\begin{table}[h]
\begin{center}
\begin{tabular}{c|cccccccc}
Bin index &1&2&3&4&5&6&7&8\\
  \hline
1&1.00&&&&&&&\\
2&0.93&1.00&&&&&&\\
3&0.84&0.80&1.00&&&&&\\
4&0.95&0.94&0.84&1.00&&&&\\
5&0.95&0.93&0.87&0.99&1.00&&&\\
6&0.74&0.79&0.70&0.86&0.85&1.00&&\\
7&0.87&0.86&0.84&0.93&0.94&0.81&1.00&\\
8&0.82&0.82&0.75&0.88&0.92&0.78&0.86&1.00\\
\end{tabular}
\end{center}
\caption{Correlation coefficients of the systematic uncertainties for the differential $W^{+}$ cross-section measurement between bins of $\eta^{e}$.}
\label{tab:Wp}
\end{table}

%% file: results/Table7.tex
\begin{table}[h]
\begin{center}
\begin{tabular}{c|cccccccc}
Bin index &1&2&3&4&5&6&7&8\\
  \hline
1&1.00&&&&&&&\\
2&0.99&1.00&&&&&&\\
3&0.99&0.99&1.00&&&&&\\
4&0.98&0.99&0.99&1.00&&&&\\
5&0.98&0.97&0.98&0.99&1.00&&&\\
6&0.72&0.75&0.72&0.77&0.76&1.00&&\\
7&0.88&0.89&0.87&0.93&0.93&0.81&1.00&\\
8&0.84&0.82&0.82&0.87&0.90&0.83&0.95&1.00\\
\end{tabular}
\end{center}
\caption{Correlation coefficients of the systematic uncertainties for the differential $W^{-}$ cross-section measurement between bins of $\eta^{e}$.}
\label{tab:Wm}
\end{table}

%% file: results/Table8.tex
\begin{table}[h]
\begin{center}
\begin{tabular}{c|cccccccc}
Bin index &1&2&3&4&5&6&7&8\\
  \hline
1&0.94&0.95&0.91&0.90&0.86&0.47&0.77&0.67\\
2&0.87&0.85&0.85&0.81&0.79&0.36&0.65&0.59\\
3&0.91&0.90&0.93&0.89&0.90&0.56&0.74&0.74\\
4&0.93&0.92&0.90&0.88&0.87&0.45&0.81&0.71\\
5&0.95&0.93&0.92&0.90&0.91&0.53&0.85&0.81\\
6&0.61&0.64&0.65&0.62&0.60&0.67&0.69&0.60\\
7&0.84&0.82&0.81&0.81&0.83&0.53&0.86&0.82\\
8&0.84&0.79&0.81&0.79&0.84&0.53&0.81&0.84\\
\end{tabular}
\end{center}
\caption{Correlation coefficients of the systematic uncertainties for the differential $W^{+}$ and $W^{-}$ cross-section measurements between bins of $\eta^{e}$. The horizontal bin indices label bins of $\eta^{e}$ for electrons while vertical indices label bins for positrons.}
\label{tab:WpvWm}
\end{table}

%% file: LHCb_Authorship_flat_07-Jun-2016.tex
\centerline{\large\bf LHCb collaboration}
\begin{flushleft}
\small
R.~Aaij$^{40}$,
B.~Adeva$^{39}$,
M.~Adinolfi$^{48}$,
Z.~Ajaltouni$^{5}$,
S.~Akar$^{6}$,
J.~Albrecht$^{10}$,
F.~Alessio$^{40}$,
M.~Alexander$^{53}$,
S.~Ali$^{43}$,
G.~Alkhazov$^{31}$
% ...

{\footnotesize \it
% ...
$ ^{5}$Clermont Universit{\'e}, Universit{\'e} Blaise Pascal, CNRS/IN2P3, LPC, Clermont-Ferrand, France\\
$ ^{6}$CPPM, Aix-Marseille Universit{\'e}, CNRS/IN2P3, Marseille, France\\
% ...
$ ^{10}$Fakult{\"a}t Physik, Technische Universit{\"a}t Dortmund, Dortmund, Germany\\
% ...
$ ^{31}$Petersburg Nuclear Physics Institute (PNPI), Gatchina, Russia\\
% ...
$ ^{39}$Universidad de Santiago de Compostela, Santiago de Compostela, Spain\\
$ ^{40}$European Organization for Nuclear Research (CERN), Geneva, Switzerland\\
% ...
$ ^{43}$Nikhef National Institute for Subatomic Physics, Amsterdam, The Netherlands\\
% ...
$ ^{48}$H.H. Wills Physics Laboratory, University of Bristol, Bristol, United Kingdom\\
% ...
$ ^{53}$School of Physics and Astronomy, University of Glasgow, Glasgow, United Kingdom\\
% ...
}
\end{flushleft}

%% file: main.bbl
\ifx\mcitethebibliography\mciteundefinedmacro
\PackageError{LHCb.bst}{mciteplus.sty has not been loaded}
{This bibstyle requires the use of the mciteplus package.}\fi
\providecommand{\href}[2]{#2}
\begin{mcitethebibliography}{10}
\mciteSetBstSublistMode{n}
\mciteSetBstMaxWidthForm{subitem}{\alph{mcitesubitemcount})}
\mciteSetBstSublistLabelBeginEnd{\mcitemaxwidthsubitemform\space}
{\relax}{\relax}

\bibitem{PDF4LHC}
J.~Butterworth {\em et~al.},
  \ifthenelse{\boolean{articletitles}}{\emph{{PDF4LHC recommendations for LHC
  Run II}}, }{}\href{http://dx.doi.org/10.1088/0954-3899/43/2/023001}{J.\
  Phys.\  \textbf{G43} (2016) 023001},
  \href{http://arxiv.org/abs/1510.03865}{{\tt arXiv:1510.03865}}\relax
\mciteBstWouldAddEndPuncttrue
\mciteSetBstMidEndSepPunct{\mcitedefaultmidpunct}
{\mcitedefaultendpunct}{\mcitedefaultseppunct}\relax
\EndOfBibitem
\bibitem{WZATLAS}
ATLAS collaboration, G.~Aad {\em et~al.},
  \ifthenelse{\boolean{articletitles}}{\emph{{Measurement of the inclusive
  $W^\pm$ and $Z/\gamma^*$ cross sections in the electron and muon decay
  channels in $pp$ collisions at \mbox{$\sqrt{s}=7$ TeV} with the ATLAS
  detector}}, }{}\href{http://dx.doi.org/10.1103/PhysRevD.85.072004}{Phys.\
  Rev.\  \textbf{D85} (2012) 072004},
  \href{http://arxiv.org/abs/1109.5141}{{\tt arXiv:1109.5141}}\relax
\mciteBstWouldAddEndPuncttrue
\mciteSetBstMidEndSepPunct{\mcitedefaultmidpunct}
{\mcitedefaultendpunct}{\mcitedefaultseppunct}\relax
\EndOfBibitem
\bibitem{ATLASDY8TeV}
ATLAS collaboration, G.~Aad {\em et~al.},
  \ifthenelse{\boolean{articletitles}}{\emph{{Measurement of the
  double-differential high-mass Drell$\,$\mydash{Yan} cross section in $pp$
  collisions at $\sqrt{s}$ = 8 TeV with the ATLAS detector}},
  }{}\href{http://dx.doi.org/10.1007/JHEP08(2016)009}{JHEP \textbf{08} (2016)
  009}, \href{http://arxiv.org/abs/1606.01736}{{\tt arXiv:1606.01736}}\relax
\mciteBstWouldAddEndPuncttrue
\mciteSetBstMidEndSepPunct{\mcitedefaultmidpunct}
{\mcitedefaultendpunct}{\mcitedefaultseppunct}\relax
\EndOfBibitem
\bibitem{ATLAS13TeV}
ATLAS collaboration, G.~Aad {\em et~al.},
  \ifthenelse{\boolean{articletitles}}{\emph{{Measurement
  of$\phantom{0}$$W^{\pm}$ and $Z$-boson production cross sections in $pp$
  collisions at $\sqrt{s}=13$ TeV with the ATLAS detector}},
  }{}\href{http://dx.doi.org/10.1016/j.physletb.2016.06.023}{Phys.\ Lett.\
  \textbf{B759} (2016) 601}, \href{http://arxiv.org/abs/1603.09222}{{\tt
  arXiv:1603.09222}}\relax
\mciteBstWouldAddEndPuncttrue
\mciteSetBstMidEndSepPunct{\mcitedefaultmidpunct}
{\mcitedefaultendpunct}{\mcitedefaultseppunct}\relax
\EndOfBibitem
\bibitem{CMSWZpT8TeV}
CMS collaboration, V.~Khachatryan {\em et~al.},
  \ifthenelse{\boolean{articletitles}}{\emph{{Measurement of the transverse
  momentum spectra of weak vector bosons produced in proton\mydash{proton}
  collisions at $\sqrt{s}=8$ TeV}},
  }{}\href{http://arxiv.org/abs/1606.05864}{{\tt arXiv:1606.05864}}, submitted
  to JHEP\relax
\mciteBstWouldAddEndPuncttrue
\mciteSetBstMidEndSepPunct{\mcitedefaultmidpunct}
{\mcitedefaultendpunct}{\mcitedefaultseppunct}\relax
\EndOfBibitem
\bibitem{CMSWasymmetry8TeV}
CMS collaboration, V.~Khachatryan {\em et~al.},
  \ifthenelse{\boolean{articletitles}}{\emph{{Measurement of the differential
  cross section and charge asymmetry for inclusive $pp$ \to W + X production at
  $\sqrt{s}=8$ TeV}},
  }{}\href{http://dx.doi.org/10.1140/epjc/s10052-016-4293-4}{Eur.\ Phys.\ J.\
  \textbf{C76} (2016), no.~8 469}, \href{http://arxiv.org/abs/1603.01803}{{\tt
  arXiv:1603.01803}}\relax
\mciteBstWouldAddEndPuncttrue
\mciteSetBstMidEndSepPunct{\mcitedefaultmidpunct}
{\mcitedefaultendpunct}{\mcitedefaultseppunct}\relax
\EndOfBibitem
\bibitem{CMSWZ8TeV}
CMS collaboration, S.~Chatrchyan {\em et~al.},
  \ifthenelse{\boolean{articletitles}}{\emph{{Measurement of inclusive W and Z
  boson production cross sections in $pp$ collisions at $\sqrt{s}$ = 8 TeV}},
  }{}\href{http://dx.doi.org/10.1103/PhysRevLett.112.191802}{Phys.\ Rev.\
  Lett.\  \textbf{112} (2014) 191802},
  \href{http://arxiv.org/abs/1402.0923}{{\tt arXiv:1402.0923}}\relax
\mciteBstWouldAddEndPuncttrue
\mciteSetBstMidEndSepPunct{\mcitedefaultmidpunct}
{\mcitedefaultendpunct}{\mcitedefaultseppunct}\relax
\EndOfBibitem
\bibitem{WmunuLHCb}
LHCb collaboration, R.~Aaij {\em et~al.},
  \ifthenelse{\boolean{articletitles}}{\emph{{Measurement of the forward $W$
  boson production cross-section in $pp$ collisions at $\sqrt{s}=7$ TeV}},
  }{}\href{http://dx.doi.org/10.1007/JHEP12(2014)079}{JHEP \textbf{12} (2014)
  079}, \href{http://arxiv.org/abs/1408.4354}{{\tt arXiv:1408.4354}}\relax
\mciteBstWouldAddEndPuncttrue
\mciteSetBstMidEndSepPunct{\mcitedefaultmidpunct}
{\mcitedefaultendpunct}{\mcitedefaultseppunct}\relax
\EndOfBibitem
\bibitem{WZLHCb8TeV}
LHCb collaboration, R.~Aaij {\em et~al.},
  \ifthenelse{\boolean{articletitles}}{\emph{{Measurement of forward W and Z
  boson production in $pp$ collisions at $ \sqrt{s}=8 $ TeV}},
  }{}\href{http://dx.doi.org/10.1007/JHEP01(2016)155}{JHEP \textbf{01} (2016)
  155}, \href{http://arxiv.org/abs/1511.08039}{{\tt arXiv:1511.08039}}\relax
\mciteBstWouldAddEndPuncttrue
\mciteSetBstMidEndSepPunct{\mcitedefaultmidpunct}
{\mcitedefaultendpunct}{\mcitedefaultseppunct}\relax
\EndOfBibitem
\bibitem{ZmumuLHCb}
LHCb collaboration, R.~Aaij {\em et~al.},
  \ifthenelse{\boolean{articletitles}}{\emph{{Measurement of the forward $Z$
  boson production cross-section in $pp$ collisions at $\sqrt{s}=7$ TeV}},
  }{}\href{http://dx.doi.org/10.1007/JHEP08(2015)039}{JHEP \textbf{08} (2015)
  039}, \href{http://arxiv.org/abs/1505.07024}{{\tt arXiv:1505.07024}}\relax
\mciteBstWouldAddEndPuncttrue
\mciteSetBstMidEndSepPunct{\mcitedefaultmidpunct}
{\mcitedefaultendpunct}{\mcitedefaultseppunct}\relax
\EndOfBibitem
\bibitem{ZeeLHCb}
LHCb collaboration, R.~Aaij {\em et~al.},
  \ifthenelse{\boolean{articletitles}}{\emph{{Measurement of the cross-section
  for $Z \to e^+e^-$ production in $pp$ collisions at $\sqrt{s}=7$ TeV}},
  }{}\href{http://dx.doi.org/10.1007/JHEP02(2013)106}{JHEP \textbf{02} (2013)
  106}, \href{http://arxiv.org/abs/1212.4620}{{\tt arXiv:1212.4620}}\relax
\mciteBstWouldAddEndPuncttrue
\mciteSetBstMidEndSepPunct{\mcitedefaultmidpunct}
{\mcitedefaultendpunct}{\mcitedefaultseppunct}\relax
\EndOfBibitem
\bibitem{ZeeLHCb8TeV}
LHCb collaboration, R.~Aaij {\em et~al.},
  \ifthenelse{\boolean{articletitles}}{\emph{{Measurement of forward $Z \to
  e^+e^-$ production in $pp$ collisions at $\sqrt{s}=8$ TeV}},
  }{}\href{http://dx.doi.org/10.1007/JHEP05(2015)109}{JHEP \textbf{05} (2015)
  109}, \href{http://arxiv.org/abs/1503.00963}{{\tt arXiv:1503.00963}}\relax
\mciteBstWouldAddEndPuncttrue
\mciteSetBstMidEndSepPunct{\mcitedefaultmidpunct}
{\mcitedefaultendpunct}{\mcitedefaultseppunct}\relax
\EndOfBibitem
\bibitem{ZtautauLHCb}
LHCb collaboration, R.~Aaij {\em et~al.},
  \ifthenelse{\boolean{articletitles}}{\emph{{A study of the $Z$ production
  cross-section in $pp$ collisions at $\sqrt{s} = 7$ TeV using tau final
  states}}, }{}\href{http://dx.doi.org/10.1007/JHEP01(2013)111}{JHEP
  \textbf{01} (2013) 111}, \href{http://arxiv.org/abs/1210.6289}{{\tt
  arXiv:1210.6289}}\relax
\mciteBstWouldAddEndPuncttrue
\mciteSetBstMidEndSepPunct{\mcitedefaultmidpunct}
{\mcitedefaultendpunct}{\mcitedefaultseppunct}\relax
\EndOfBibitem
\bibitem{Z13TeV}
LHCb collaboration, R.~Aaij {\em et~al.},
  \ifthenelse{\boolean{articletitles}}{\emph{{Measurement of the forward Z
  boson production cross-section in pp collisions at $\sqrt{s} = 13$ TeV}},
  }{}\href{http://dx.doi.org/doi:10.1007/JHEP09(2016)136}{JHEP \textbf{09}
  (2016) 136}, \href{http://arxiv.org/abs/1607.06495}{{\tt
  arXiv:1607.06495}}\relax
\mciteBstWouldAddEndPuncttrue
\mciteSetBstMidEndSepPunct{\mcitedefaultmidpunct}
{\mcitedefaultendpunct}{\mcitedefaultseppunct}\relax
\EndOfBibitem
\bibitem{FEWZ}
R.~Gavin, Y.~Li, F.~Petriello, and S.~Quackenbush,
  \ifthenelse{\boolean{articletitles}}{\emph{{FEWZ 2.0: A code for hadronic Z
  production at next-to-next-to-leading order}},
  }{}\href{http://dx.doi.org/10.1016/j.cpc.2011.06.008}{Comput.\ Phys.\
  Commun.\  \textbf{182} (2011) 2388},
  \href{http://arxiv.org/abs/1011.3540}{{\tt arXiv:1011.3540}}\relax
\mciteBstWouldAddEndPuncttrue
\mciteSetBstMidEndSepPunct{\mcitedefaultmidpunct}
{\mcitedefaultendpunct}{\mcitedefaultseppunct}\relax
\EndOfBibitem
\bibitem{FEWZ2}
Y.~Li and F.~Petriello, \ifthenelse{\boolean{articletitles}}{\emph{{Combining
  QCD and electroweak corrections to dilepton production in FEWZ}},
  }{}\href{http://dx.doi.org/10.1103/PhysRevD.86.094034}{Phys.\ Rev.\
  \textbf{D86} (2012) 094034}, \href{http://arxiv.org/abs/1208.5967}{{\tt
  arXiv:1208.5967}}\relax
\mciteBstWouldAddEndPuncttrue
\mciteSetBstMidEndSepPunct{\mcitedefaultmidpunct}
{\mcitedefaultendpunct}{\mcitedefaultseppunct}\relax
\EndOfBibitem
\bibitem{Alves:2008zz}
LHCb collaboration, A.~A. Alves~Jr.\ {\em et~al.},
  \ifthenelse{\boolean{articletitles}}{\emph{{The \lhcb detector at the LHC}},
  }{}\href{http://dx.doi.org/10.1088/1748-0221/3/08/S08005}{JINST \textbf{3}
  (2008) S08005}\relax
\mciteBstWouldAddEndPuncttrue
\mciteSetBstMidEndSepPunct{\mcitedefaultmidpunct}
{\mcitedefaultendpunct}{\mcitedefaultseppunct}\relax
\EndOfBibitem
\bibitem{LHCb-DP-2014-002}
{LHCb collaboration, R.\ Aaij $et\ al$},
  \ifthenelse{\boolean{articletitles}}{\emph{{LHCb detector performance}},
  }{}\href{http://dx.doi.org/10.1142/S0217751X15300227}{Int.\ J.\ Mod.\ Phys.\
  \textbf{A30} (2015) 1530022}, \href{http://arxiv.org/abs/1412.6352}{{\tt
  arXiv:1412.6352}}\relax
\mciteBstWouldAddEndPuncttrue
\mciteSetBstMidEndSepPunct{\mcitedefaultmidpunct}
{\mcitedefaultendpunct}{\mcitedefaultseppunct}\relax
\EndOfBibitem
\bibitem{Sjostrand:2007gs}
T.~Sj\"{o}strand, S.~Mrenna, and P.~Skands,
  \ifthenelse{\boolean{articletitles}}{\emph{{A brief introduction to PYTHIA
  8.1}}, }{}\href{http://dx.doi.org/10.1016/j.cpc.2008.01.036}{Comput.\ Phys.\
  Commun.\  \textbf{178} (2008) 852},
  \href{http://arxiv.org/abs/0710.3820}{{\tt arXiv:0710.3820}}\relax
\mciteBstWouldAddEndPuncttrue
\mciteSetBstMidEndSepPunct{\mcitedefaultmidpunct}
{\mcitedefaultendpunct}{\mcitedefaultseppunct}\relax
\EndOfBibitem
\bibitem{Sjostrand:2006za}
T.~Sj\"{o}strand, S.~Mrenna, and P.~Skands,
  \ifthenelse{\boolean{articletitles}}{\emph{{PYTHIA 6.4 physics and manual}},
  }{}\href{http://dx.doi.org/10.1088/1126-6708/2006/05/026}{JHEP \textbf{05}
  (2006) 026}, \href{http://arxiv.org/abs/hep-ph/0603175}{{\tt
  arXiv:hep-ph/0603175}}\relax
\mciteBstWouldAddEndPuncttrue
\mciteSetBstMidEndSepPunct{\mcitedefaultmidpunct}
{\mcitedefaultendpunct}{\mcitedefaultseppunct}\relax
\EndOfBibitem
\bibitem{LHCb-PROC-2010-056}
I.~Belyaev {\em et~al.}, \ifthenelse{\boolean{articletitles}}{\emph{{Handling
  of the generation of primary events in Gauss, the LHCb simulation
  framework}}, }{}\href{http://dx.doi.org/10.1088/1742-6596/331/3/032047}{{J.\
  Phys.\ Conf.\ Ser.\ } \textbf{331} (2011) 032047}\relax
\mciteBstWouldAddEndPuncttrue
\mciteSetBstMidEndSepPunct{\mcitedefaultmidpunct}
{\mcitedefaultendpunct}{\mcitedefaultseppunct}\relax
\EndOfBibitem
\bibitem{Allison:2006ve}
Geant4 collaboration, J.~Allison {\em et~al.},
  \ifthenelse{\boolean{articletitles}}{\emph{{Geant4 developments and
  applications}}, }{}\href{http://dx.doi.org/10.1109/TNS.2006.869826}{IEEE
  Trans.\ Nucl.\ Sci.\  \textbf{53} (2006) 270}\relax
\mciteBstWouldAddEndPuncttrue
\mciteSetBstMidEndSepPunct{\mcitedefaultmidpunct}
{\mcitedefaultendpunct}{\mcitedefaultseppunct}\relax
\EndOfBibitem
\bibitem{Agostinelli:2002hh}
Geant4 collaboration, S.~Agostinelli {\em et~al.},
  \ifthenelse{\boolean{articletitles}}{\emph{{Geant4: A simulation toolkit}},
  }{}\href{http://dx.doi.org/10.1016/S0168-9002(03)01368-8}{Nucl.\ Instrum.\
  Meth.\  \textbf{A506} (2003) 250}\relax
\mciteBstWouldAddEndPuncttrue
\mciteSetBstMidEndSepPunct{\mcitedefaultmidpunct}
{\mcitedefaultendpunct}{\mcitedefaultseppunct}\relax
\EndOfBibitem
\bibitem{LHCb-PROC-2011-006}
M.~Clemencic {\em et~al.}, \ifthenelse{\boolean{articletitles}}{\emph{{The
  \lhcb simulation application, Gauss: Design, evolution and experience}},
  }{}\href{http://dx.doi.org/10.1088/1742-6596/331/3/032023}{{J.\ Phys.\ Conf.\
  Ser.\ } \textbf{331} (2011) 032023}\relax
\mciteBstWouldAddEndPuncttrue
\mciteSetBstMidEndSepPunct{\mcitedefaultmidpunct}
{\mcitedefaultendpunct}{\mcitedefaultseppunct}\relax
\EndOfBibitem
\bibitem{CTEQ6L1}
P.~M. Nadolsky {\em et~al.},
  \ifthenelse{\boolean{articletitles}}{\emph{{Implications of CTEQ global
  analysis for collider observables}},
  }{}\href{http://dx.doi.org/10.1103/PhysRevD.78.013004}{Phys.\ Rev.\
  \textbf{D78} (2008) 013004}, \href{http://arxiv.org/abs/0802.0007}{{\tt
  arXiv:0802.0007}}\relax
\mciteBstWouldAddEndPuncttrue
\mciteSetBstMidEndSepPunct{\mcitedefaultmidpunct}
{\mcitedefaultendpunct}{\mcitedefaultseppunct}\relax
\EndOfBibitem
\bibitem{pythiaFSR}
T.~Sj{\"o}strand and P.~Z. Skands,
  \ifthenelse{\boolean{articletitles}}{\emph{{Transverse-momentum-ordered
  showers and interleaved multiple interactions}},
  }{}\href{http://dx.doi.org/10.1140/epjc/s2004-02084-y}{Eur.\ Phys.\ J.\
  \textbf{C39} (2005) 129}, \href{http://arxiv.org/abs/hep-ph/0408302}{{\tt
  arXiv:hep-ph/0408302}}\relax
\mciteBstWouldAddEndPuncttrue
\mciteSetBstMidEndSepPunct{\mcitedefaultmidpunct}
{\mcitedefaultendpunct}{\mcitedefaultseppunct}\relax
\EndOfBibitem
\bibitem{fitting}
R.~J. Barlow and C.~Beeston,
  \ifthenelse{\boolean{articletitles}}{\emph{{Fitting using finite Monte Carlo
  samples}}, }{}\href{http://dx.doi.org/10.1016/0010-4655(93)90005-W}{Comput.\
  Phys.\ Commun.\  \textbf{77} (1993) 219}\relax
\mciteBstWouldAddEndPuncttrue
\mciteSetBstMidEndSepPunct{\mcitedefaultmidpunct}
{\mcitedefaultendpunct}{\mcitedefaultseppunct}\relax
\EndOfBibitem
\bibitem{PDG}
Particle Data Group, K.~A. Olive {\em et~al.},
  \ifthenelse{\boolean{articletitles}}{\emph{{\href{http://pdg.lbl.gov/}{Review
  of particle physics}}},
  }{}\href{http://dx.doi.org/10.1088/1674-1137/38/9/090001}{Chin.\ Phys.\
  \textbf{C38} (2014) 090001}, {and 2015 update}\relax
\mciteBstWouldAddEndPuncttrue
\mciteSetBstMidEndSepPunct{\mcitedefaultmidpunct}
{\mcitedefaultendpunct}{\mcitedefaultseppunct}\relax
\EndOfBibitem
\bibitem{MCFM}
J.~M. Campbell and R.~K. Ellis,
  \ifthenelse{\boolean{articletitles}}{\emph{{MCFM for the Tevatron and the
  LHC}}, }{}\href{http://dx.doi.org/10.1016/j.nuclphysbps.2010.08.011}{Nucl.\
  Phys.\ Proc.\ Suppl.\  \textbf{205-206} (2010) 10},
  \href{http://arxiv.org/abs/1007.3492}{{\tt arXiv:1007.3492}}\relax
\mciteBstWouldAddEndPuncttrue
\mciteSetBstMidEndSepPunct{\mcitedefaultmidpunct}
{\mcitedefaultendpunct}{\mcitedefaultseppunct}\relax
\EndOfBibitem
\bibitem{RESBOS1}
G.~A. Ladinsky and C.~P. Yuan, \ifthenelse{\boolean{articletitles}}{\emph{{The
  Nonperturbative regime in QCD resummation for gauge boson production at
  hadron colliders}},
  }{}\href{http://dx.doi.org/10.1103/PhysRevD.50.R4239}{Phys.\ Rev.\
  \textbf{D50} (1994) 4239}, \href{http://arxiv.org/abs/hep-ph/9311341}{{\tt
  arXiv:hep-ph/9311341}}\relax
\mciteBstWouldAddEndPuncttrue
\mciteSetBstMidEndSepPunct{\mcitedefaultmidpunct}
{\mcitedefaultendpunct}{\mcitedefaultseppunct}\relax
\EndOfBibitem
\bibitem{RESBOS2}
C.~Balazs and C.~P. Yuan, \ifthenelse{\boolean{articletitles}}{\emph{{Soft
  gluon effects on lepton pairs at hadron colliders}},
  }{}\href{http://dx.doi.org/10.1103/PhysRevD.56.5558}{Phys.\ Rev.\
  \textbf{D56} (1997) 5558}, \href{http://arxiv.org/abs/hep-ph/9704258}{{\tt
  arXiv:hep-ph/9704258}}\relax
\mciteBstWouldAddEndPuncttrue
\mciteSetBstMidEndSepPunct{\mcitedefaultmidpunct}
{\mcitedefaultendpunct}{\mcitedefaultseppunct}\relax
\EndOfBibitem
\bibitem{RESBOS3}
F.~Landry, R.~Brock, P.~M. Nadolsky, and C.~P. Yuan,
  \ifthenelse{\boolean{articletitles}}{\emph{{Tevatron Run-1 $Z$ boson data and
  Collins-Soper-Sterman resummation formalism}},
  }{}\href{http://dx.doi.org/10.1103/PhysRevD.67.073016}{Phys.\ Rev.\
  \textbf{D67} (2003) 073016}, \href{http://arxiv.org/abs/hep-ph/0212159}{{\tt
  arXiv:hep-ph/0212159}}\relax
\mciteBstWouldAddEndPuncttrue
\mciteSetBstMidEndSepPunct{\mcitedefaultmidpunct}
{\mcitedefaultendpunct}{\mcitedefaultseppunct}\relax
\EndOfBibitem
\bibitem{herwig}
M.~Bahr {\em et~al.}, \ifthenelse{\boolean{articletitles}}{\emph{{Herwig++
  physics and manual}},
  }{}\href{http://dx.doi.org/10.1140/epjc/s10052-008-0798-9}{Eur.\ Phys.\ J.\
  \textbf{C58} (2008) 639}, \href{http://arxiv.org/abs/0803.0883}{{\tt
  arXiv:0803.0883}}\relax
\mciteBstWouldAddEndPuncttrue
\mciteSetBstMidEndSepPunct{\mcitedefaultmidpunct}
{\mcitedefaultendpunct}{\mcitedefaultseppunct}\relax
\EndOfBibitem
\bibitem{beam}
J.~Wenninger, \ifthenelse{\boolean{articletitles}}{\emph{{Energy calibration of
  the LHC beams at 4 TeV}}, }{} Tech. Rep. \mbox{CERN-ATS-2013-040}, CERN,
  Geneva\relax
\mciteBstWouldAddEndPuncttrue
\mciteSetBstMidEndSepPunct{\mcitedefaultmidpunct}
{\mcitedefaultendpunct}{\mcitedefaultseppunct}\relax
\EndOfBibitem
\bibitem{dynnlo}
S.~Catani {\em et~al.}, \ifthenelse{\boolean{articletitles}}{\emph{{Vector
  boson production at hadron colliders: A fully exclusive QCD calculation at
  NNLO}}, }{}\href{http://dx.doi.org/10.1103/PhysRevLett.103.082001}{Phys.\
  Rev.\ Lett.\  \textbf{103} (2009) 082001},
  \href{http://arxiv.org/abs/0903.2120}{{\tt arXiv:0903.2120}}\relax
\mciteBstWouldAddEndPuncttrue
\mciteSetBstMidEndSepPunct{\mcitedefaultmidpunct}
{\mcitedefaultendpunct}{\mcitedefaultseppunct}\relax
\EndOfBibitem
\bibitem{luminosity}
LHCb collaboration, R.~Aaij {\em et~al.},
  \ifthenelse{\boolean{articletitles}}{\emph{{Precision luminosity measurements
  at LHCb}}, }{}\href{http://dx.doi.org/10.1088/1748-0221/9/12/P12005}{JINST
  \textbf{9} (2014) P12005}, \href{http://arxiv.org/abs/1410.0149}{{\tt
  arXiv:1410.0149}}\relax
\mciteBstWouldAddEndPuncttrue
\mciteSetBstMidEndSepPunct{\mcitedefaultmidpunct}
{\mcitedefaultendpunct}{\mcitedefaultseppunct}\relax
\EndOfBibitem
\bibitem{ABM}
S.~Alekhin, J.~Bl\"{u}mlein, and S.~Moch,
  \ifthenelse{\boolean{articletitles}}{\emph{{The ABM parton distributions
  tuned to LHC data}},
  }{}\href{http://dx.doi.org/10.1103/PhysRevD.89.054028}{Phys.\ Rev.\
  \textbf{D89} (2014) 054028}, \href{http://arxiv.org/abs/1310.3059}{{\tt
  arXiv:1310.3059}}\relax
\mciteBstWouldAddEndPuncttrue
\mciteSetBstMidEndSepPunct{\mcitedefaultmidpunct}
{\mcitedefaultendpunct}{\mcitedefaultseppunct}\relax
\EndOfBibitem
\bibitem{CT14}
S.~Dulat {\em et~al.}, \ifthenelse{\boolean{articletitles}}{\emph{{New parton
  distribution functions from a global analysis of quantum chromodynamics}},
  }{}\href{http://dx.doi.org/10.1103/PhysRevD.93.033006}{Phys.\ Rev.\
  \textbf{D93} (2016) 033006}, \href{http://arxiv.org/abs/1506.07443}{{\tt
  arXiv:1506.07443}}\relax
\mciteBstWouldAddEndPuncttrue
\mciteSetBstMidEndSepPunct{\mcitedefaultmidpunct}
{\mcitedefaultendpunct}{\mcitedefaultseppunct}\relax
\EndOfBibitem
\bibitem{HERAPDF}
ZEUS, H1 collaborations, A.~M. Cooper-Sarkar,
  \ifthenelse{\boolean{articletitles}}{\emph{{PDF Fits at HERA}}, }{}PoS
  \textbf{EPS-HEP2011} (2011) 320, \href{http://arxiv.org/abs/1112.2107}{{\tt
  arXiv:1112.2107}}\relax
\mciteBstWouldAddEndPuncttrue
\mciteSetBstMidEndSepPunct{\mcitedefaultmidpunct}
{\mcitedefaultendpunct}{\mcitedefaultseppunct}\relax
\EndOfBibitem
\bibitem{MMHT}
L.~A. Harland-Lang, A.~D. Martin, P.~Motylinski, and R.~S. Thorne,
  \ifthenelse{\boolean{articletitles}}{\emph{{Parton distributions in the LHC
  era: MMHT 2014 PDFs}},
  }{}\href{http://dx.doi.org/10.1140/epjc/s10052-015-3397-6}{Eur.\ Phys.\ J.\
  \textbf{C75} (2015) 204}, \href{http://arxiv.org/abs/1412.3989}{{\tt
  arXiv:1412.3989}}\relax
\mciteBstWouldAddEndPuncttrue
\mciteSetBstMidEndSepPunct{\mcitedefaultmidpunct}
{\mcitedefaultendpunct}{\mcitedefaultseppunct}\relax
\EndOfBibitem
\bibitem{MSTW}
A.~D. Martin, W.~J. Stirling, R.~S. Thorne, and G.~Watt,
  \ifthenelse{\boolean{articletitles}}{\emph{{Parton distributions for the
  LHC}}, }{}\href{http://dx.doi.org/10.1140/epjc/s10052-009-1072-5}{Eur.\
  Phys.\ J.\  \textbf{C63} (2009) 189},
  \href{http://arxiv.org/abs/0901.0002}{{\tt arXiv:0901.0002}}\relax
\mciteBstWouldAddEndPuncttrue
\mciteSetBstMidEndSepPunct{\mcitedefaultmidpunct}
{\mcitedefaultendpunct}{\mcitedefaultseppunct}\relax
\EndOfBibitem
\bibitem{NNPDF}
NNPDF collaboration, R.~D. Ball {\em et~al.},
  \ifthenelse{\boolean{articletitles}}{\emph{{Parton distributions for the LHC
  Run II}}, }{}\href{http://dx.doi.org/10.1007/JHEP04(2015)040}{JHEP
  \textbf{04} (2015) 040}, \href{http://arxiv.org/abs/1410.8849}{{\tt
  arXiv:1410.8849}}\relax
\mciteBstWouldAddEndPuncttrue
\mciteSetBstMidEndSepPunct{\mcitedefaultmidpunct}
{\mcitedefaultendpunct}{\mcitedefaultseppunct}\relax
\EndOfBibitem
\bibitem{LUCDF}
CDF collaboration, A.~Abulencia {\em et~al.},
  \ifthenelse{\boolean{articletitles}}{\emph{{Measurements of inclusive W and Z
  cross sections in p$\bar{p}$ collisions at $\sqrt{s}=1.96$ TeV}},
  }{}\href{http://dx.doi.org/10.1088/0954-3899/34/12/001}{J.\ Phys.\
  \textbf{G34} (2007) 2457}, \href{http://arxiv.org/abs/hep-ex/0508029}{{\tt
  arXiv:hep-ex/0508029}}\relax
\mciteBstWouldAddEndPuncttrue
\mciteSetBstMidEndSepPunct{\mcitedefaultmidpunct}
{\mcitedefaultendpunct}{\mcitedefaultseppunct}\relax
\EndOfBibitem
\bibitem{LEPCombo}
ALEPH, DELPHI, L3, OPAL collaborations, LEP Electroweak working group,
  S.~Schael {\em et~al.},
  \ifthenelse{\boolean{articletitles}}{\emph{{Electroweak measurements in
  electron-positron collisions at W-boson-pair energies at LEP}},
  }{}\href{http://dx.doi.org/10.1016/j.physrep.2013.07.004}{Phys.\ Rept.\
  \textbf{532} (2013) 119}, \href{http://arxiv.org/abs/1302.3415}{{\tt
  arXiv:1302.3415}}\relax
\mciteBstWouldAddEndPuncttrue
\mciteSetBstMidEndSepPunct{\mcitedefaultmidpunct}
{\mcitedefaultendpunct}{\mcitedefaultseppunct}\relax
\EndOfBibitem
\end{mcitethebibliography}
